\documentstyle[11pt,amssymb]{article}

\makeatletter
\@addtoreset{equation}{section}
\makeatother

\def\dalemb#1#2{{\vbox{\hrule height .#2pt
        \hbox{\vrule width.#2pt height#1pt \kern#1pt
                \vrule width.#2pt}
        \hrule height.#2pt}}}

\def\cC{{\cal C}}

\def\cao{{\cal O}}

\def\p{\pi}

\def\0{{\sst{(0)}}}
\def\1{{\sst{(1)}}}
\def\2{{\sst{(2)}}}
\def\3{{\sst{(3)}}}
\def\4{{\sst{(4)}}}
\def\5{{\sst{(5)}}}
\def\6{{\sst{(6)}}}
\def\7{{\sst{(7)}}}
\def\8{{\sst{(8)}}}
\def\n{{\sst{(n)}}}

\def\ep{\epsilon}
\def\td{\tilde}

\def\half{{\textstyle{1\over2}}}

\let\a=\alpha \let\b=\beta \let\g=\gamma \let\d=\delta \let\e=\epsilon
\let\z=\zeta  \let\q=\theta  \let\k=\kappa
\let\l=\lambda \let\m=\mu \let\n=\nu \let\x=\xi \let\r=\rho
\let\s=\sigma \let\t=\tau  \let\f=\phi  
  \let\D=\Delta  \let\L=\Lambda
   \let\F=\Phi 
 \let\W=\Omega   \let\G=\Gamma
\let\la=\label  
  
\def\nn{\nonumber} \def\bd{\begin{document}} \def\ed{\end{document}}
\def\ds{\documentstyle} \let\fr=\frac \let\bl=\bigl \let\br=\bigr
\let\Br=\Bigr \let\Bl=\Bigl
\let\bm=\bibitem
\let\na=\nabla
\let\pa=\partial \let\ov=\overline
\newcommand{\be}{\begin{equation}}
\newcommand{\ee}{\end{equation}}
\def\ba{\begin{array}}
\def\ea{\end{array}}
\def\ft#1#2{{\textstyle{{\scriptstyle #1}\over {\scriptstyle #2}}}}
\def\fft#1#2{{#1 \over #2}}
\def\del{\partial}
\def\sst#1{{\scriptscriptstyle #1}}
 \def\oneone{\rlap 1\mkern4mu{\rm l}}
\def\ie{{\it i.e.\ }}
\def\via{{\it via}}
\def\semi{{\ltimes}}
\def\str{{\rm str}}
\def\Dm{{{D_{\sst{max}}}}}
\def\vac{ \left | 0 \right \rangle }
\def\kvac{ \left | k \right \rangle }

\def\sp{\; \; \;}

\def\bol{ \left | B (p^+) \right \rangle}
\def\bo1{ \left | B^0 (p^+) \right \rangle}

\def\bolt{ \left | B (p^+) \right \rangle_{\t}}

\def\boxl{ \left | B (x^-) \right \rangle}

\def\<{ \langle }
\def\>{ \rangle }

\newcommand{\hsp}{\hspace{0.5cm}}

\newcommand{\ho}[1]{$\, ^{#1}$}
\newcommand{\hoch}[1]{$\, ^{#1}$}
\newcommand{\bea}{\begin{eqnarray}}
\newcommand{\eea}{\end{eqnarray}}
\newcommand{\ra}{\rightarrow}
\newcommand{\lra}{\longrightarrow}
\newcommand{\Lra}{\Leftrightarrow}
\newcommand{\ap}{\alpha^\prime}
\newcommand{\bp}{\tilde \beta^\prime}
\newcommand{\tr}{{\rm tr} }
\newcommand{\Tr}{{\rm Tr} }
\newcommand{\NP}{Nucl. Phys. }

\newcommand{\ams}{{\it Institute for Theoretical Physics,
University of Amsterdam, \\
Valckenierstraat 65, 1018XE Amsterdam, The Netherlands} \\
{\tt skenderi,taylor@science.uva.nl}}
\newcommand{\auth}{\large Kostas Skenderis and Marika Taylor}

\begin{document}
\begin{flushright}
\hfill{\bf hep-th/0603016}\\
\hfill{ITFA-2006-04}
\end{flushright}

\vspace{30pt}

\begin{center}

{\Large \bf Kaluza-Klein Holography}

\vspace{20pt}

\auth

\vspace{15pt}

{\ams}

\vspace{30pt}

\underline{ABSTRACT}
\vspace{10pt}
\end{center}

We construct a holographic map between asymptotically 
$AdS_5 \times S^5$ solutions of $10d$ supergravity
and vacuum expectation values of gauge invariant 
operators of the dual QFT. The ingredients that 
enter in the construction are 
(i) gauge invariant variables so that the KK reduction
is independent of any choice of gauge fixing; (ii)
the non-linear KK reduction map from 10 to 5 dimensions
(constructed perturbatively in the number of fields);
(iii) application of holographic renormalization. 
A non-trivial role in the last step
is played by extremal couplings. This map 
allows one to reliably compute vevs of operators dual to any
KK fields. As an application we consider a Coulomb branch 
solution and compute the first two non-trivial vevs, 
involving operators of dimension 2 and 4, and reproduce 
the field theory result, in agreement with non-renormalization theorems. 
This constitutes
the first {\it quantitative} test of the gravity/gauge theory duality
away from the conformal point involving a vev of an operator 
dual to a KK field (which is not one of the gauged supergravity
fields).
 
\noindent

\pagebreak
\setcounter{page}{1}

\tableofcontents
\addtocontents{toc}{\protect\setcounter{tocdepth}{2}}
\pagebreak

\section{Introduction}

Gravity/gauge theory dualities relate string theory on spacetimes
that asymptote to $AdS_m \times X$, where $X$ is a compact manifold
and gauge theory residing on the conformal boundary of the 
$AdS$ part of the geometry. In the initial work \cite{Maldacena:1997re}
the dual 
theory was a conformal field theory (CFT) and the bulk spacetime 
$AdS_m \times X$ (rather than asymptotic to it), but it was soon
recognized that the duality can be extended to describe 
quantum field theories that can be obtained from the CFT by 
either adding new terms in the action or 
considering vacua where the conformal symmetry 
is spontaneously broken. Both of these cases are described 
gravitationally by a solution that is asymptotic to $AdS_m \times X$.

Despite much work however basic questions still remain. 
One such question that will be the subject of this 
paper is: \\
{\it Given a ten dimensional solution that is 
asymptotic to $AdS_m \times X$ how does one compute 
the vacuum expectation values of gauge invariant 
operators?}\\
Roughly speaking vevs of chiral primary operators
should appear in the radial expansion of the bulk 
solution. However making this
precise proves to be a lot more subtle that one might
have anticipated, and even qualitative features are 
not reproduced correctly by naive methods. 
The answer to this question should follow from the 
basic AdS/CFT dictionary \cite{GKP,W}. This is 
indeed the case but in order to implement the
idea one has to sharpen existing methods and
overcome several technical issues. 

To illustrate the issues involved it is instructive
to consider a simple example where the physics of the solution
is well understood. A class of such examples is provided by
multicenter D3-brane solutions, which
in the near-horizon limit correspond to the Coulomb branch 
of ${\cal N}=4$ SYM \cite{Kraus:1998hv}. These examples are particularly
interesting because the QFT vevs are protected by a non-renormalization
theorem, and the gravitational results must therefore agree exactly with 
those computed at weak coupling. The metric is of the well-known form
\be \label{sol}
ds^2 = H(x_\perp) ^{-1/2} d x_{||}^2 + H(x_\perp)^{1/2} d x_\perp^2
\ee
where $H$ is a harmonic function in transverse directions.
For a distribution $\s(\vec{y})$ of D3 branes, the harmonic 
function reads
\be \label{H_CB}
H(x_\perp) = \int d^6 y \frac{\s(\vec{y})}{|\vec{x}_\perp - \vec{y}|^4}
=\frac{Q_0}{r^4}\left(1+ \sum_{k=1}^\infty \frac{Q_k Y^k}{r^{k}}\right) 
\ee
where in the last equality we expanded in $r^2=|x_\perp|^2$, $Y^k$
are spherical harmonics and $Q_k$ are numerical constants that depend
on the distribution $\s(\vec{y})$. Inserting this in (\ref{sol})
and expanding in $r$ results in a metric that is asymptotically 
$AdS_5 \times S^5$. The QFT vevs should be encoded in the asymptotics
and the purpose of this work is to show how to unambiguously
extract this information.

The solutions under discussion are special in that they are
uniquely determined in terms of a harmonic function. Furthermore,
the spherical harmonics appearing in (\ref{H_CB}) are in 1-1 correspondence 
with the chiral primary operators of ${\cal N}=4$ 
SYM and the radial power at which 
they appear is the correct power for their coefficients to correspond 
to the vevs of the dual operator. This led \cite{Klebanov:1999tb} 
to propose that these coefficients are proportional to vevs of the 
corresponding operators. Although this is a well motivated proposal, 
it is not clear how one would generalize it
to the general case where the solution is not determined by a harmonic 
function. Even for the  case at hand
there are several open questions. For instance, inserting
the harmonic function in the metric leads to terms involving 
powers of spherical harmonics whose meaning is not clear and 
in general there is also a dependence on the radial coordinate used to perform
the asymptotic expansion. The simplifications special to such cases
(i.e. when the solution is determined by a 
harmonic function) will be discussed in a separate 
publication \cite{ST}. 
In this paper we strive for generality, so our starting point will be general 
asymptotically $AdS_5 \times S^5$ metrics and we will only 
use the CB solutions in order to illustrate the general procedure.
 
Recall that the basic dictionary of the gravity/gauge theory duality
\cite{GKP,W}
states that (i) there is a bulk field corresponding to 
each gauge invariant operator
and (ii) the string partition function with bulk fields satisfying
appropriate boundary  conditions
is equal to the generating functional of QFT correlators
with the boundary conditions playing the role of sources.
In particular, one could compute vevs (in the large $N$ and large 
't Hooft coupling limit) by differentiating 
the supergravity on-shell action once w.r.t. sources.
In implementing this procedure however one finds several obstacles.

First, the relation in (ii) should be understood as a ``bare relation'' as 
both sides divergence.
To make the procedure well-defined one must renormalize.
This is a standard procedure on the field theory side.
On the gravitational side, the corresponding procedure, denoted
holographic renormalization, was developed in a series 
of papers \cite{dHSS,BFS1,BFS2,Papadimitriou:2004rz} 
(see also \cite{Henningson:1998gx}-\cite{Papadimitriou:2004ap}
for related work and \cite{Skenderis:2002wp} for a review)\footnote{
The starting point in the analysis in these 
papers was the lower dimensional AdS gravity obtained by reducing
the original theory over the compact space $X$. 
A discussion that starts from higher dimensions can be found in 
\cite{Taylor-Robinson:2001pp,Taylor-Robinson:2001fe}.}.
After renormalization
is done, the one point functions can be computed in all generality.
The answer relates the one point function to certain
coefficients in the asymptotic expansion of the bulk fields.
So given any solution of 5d gravity coupled to matter one 
could read off the vevs of the dual operators by looking at the
asymptotics. 

We would like to emphasize that the procedure of 
renormalization is essential for correctly extracting the vevs.
To give an example where a naive prescription fails, consider 
the case of the CB solution corresponding to a distribution
of D3 branes on a disc of radius $l$.
The 5d metric obtained by reducing the 10d solution 
over the sphere has the following asymptotics,
\be \label{5d_CB}
ds^2 = \frac{d \hat{z}^2}{\hat{z}^2} 
+ \frac{1}{\hat{z}^2}\left(1-\frac{l^4}{18} \hat{z}^4 
+ \cao(\hat{z}^6)\right) dx^i dx^i
\ee
A naive prescription for reading off vevs that is often quoted in the 
literature is that the vev of an operator can be 
obtained, up to a (non-zero) numerical constant, from the normalizable  
mode of the corresponding bulk field. 
The bulk metric is dual to the stress energy tensor so one would be tempted to 
identify the coefficient of the $\hat{z}^4$ term with the vev of the stress
energy tensor. This is clearly incorrect since that would imply 
non-zero vacuum energy for the dual theory but the solution is supersymmetric
so the vacuum energy should be equal to zero. Indeed the 1-point function
extracted using holographic renormalization \cite{BFS1,BFS2} 
contains additional terms (see (\ref{tij}) below) and taking those
into account one finds the expected result, zero. Such subtleties are
 present in all
cases, including that of scalar fields. For example, for deformation
flows the vevs of all operators should be equal to zero, but 
there are examples where the above naive prescription leads to non-zero values.
Again the correct 1-point functions include additional
terms so that the total result is zero (see \cite{BFS1,BFS2,Karch:2005ms}).

An analysis that starts from the lower dimensional gauged supergravity 
is sufficient if one is only interested in computing vevs for operators 
dual to fields of the gauged supergravity. 
There is however an infinity of other
(half supersymmetric) gauge invariant operators which would then 
{\it ab initio} be excluded from the analysis. These operators are dual 
to massive KK fields. The map between KK fields and gauge 
invariant operators was worked out in \cite{W} (and subsequent papers) 
using the computation of the KK spectrum of $AdS_5 \times S^5$ in 
\cite{Kim:1985ez}. In this paper a {\it linearized} analysis
around $AdS_5 \times S^5$ was performed.
This analysis provides an explicit map 
(in a {\it specific gauge}) between linearized 
solutions of the ten dimensional equations  of motion and linearized solutions
of the dimensionally reduced five dimensional equations. To compute the 
vevs however we need to know the map at the non-linear level. To 
illustrate this, let $s^k$ be ten-dimensional fields and let $S^k$ 
be the corresponding five dimensional fields. 
In general the reduction map will be non-linear and takes the form
\be
S^k = s^k + \sum_{lm} \left(J_{klm} s^l s^m 
+ L_{klm} D_\mu s^l D^\mu s^m  + \cao[s]^3\right)
\ee
where $J_{klm}$ and $L_{klm}$ are numerical coefficients and we
retain only terms quadratic in the fields. (We also suppress
contributions on the right hand side from other scalar and non-scalar
fields since they are not necessary to illustrate our point.)
If $S^k$ is dual to an operator of dimension $k$ then we would need
to extract the coefficient of order $z^k$ to determine the operator's
vev. Clearly, quadratic
terms with $l+m=k$ will also contribute at the same order
and therefore such non-linear terms in the KK map cannot be
ignored. Similarly cubic and higher order in fluctuation
terms that are of order $z^k$ will also
contribute, along with non-linear contributions involving other
supergravity fields.

So to read off the vevs we need to understand the KK reduction map 
at the non-linear level. However, if we are interested in the vev of 
an operator of a given dimension, only certain non-linear terms need
to be computed, namely  the ones that could possibly contribute to the vev.
For instance, if we are interested in computing the vev of an operator
of dimension 4, we would only need to keep terms quadratic in the fields
dual to operators of dimension 2. This in effect truncates the 
reduction to a finite number of fields. This should be contrasted with the 
issue of consistent truncation. When the latter is possible 
one keeps only the ``massless'' KK modes in the reduction.
In our case we keep massive KK fields as well.
However, only a finite subset of them contribute to the asymptotics
up to a given order.

Another issue is that of gauge fixing. The analysis in 
\cite{Kim:1985ez} was done in the de Donder gauge. Generically
however a given supergravity solution will not be in this gauge
and finding the coordinate transformation that would bring the solution
to this gauge is a complicated task. To deal with this
issue we will instead develop a ``gauge invariant KK reduction''.
Instead of fixing the gauge, we combine the fields in gauge 
invariant combinations. This can be done systematically in 
an expansion in the number of fields. Having worked out these
combinations, one can relax the de Donder gauge condition by simply replacing 
every field by its gauge invariant generalization in all results
obtained in a specific gauge.

To summarize, we argue that in order to compute the vevs we need 
to obtain the non-linear KK map in terms of gauge invariant
variables to appropriate order in the number of fields.
This procedure results in five dimensional field equations
and an explicit map between 10d solutions and  solutions of these
5d equations. 
The 5d equations can be integrated into a 5d action
and from here one can obtain the 1-point functions following
the procedure of holographic renormalization. 
There is however an additional subtlety. In some cases
the five dimensional equations contain no couplings between 
certain fields but boundary interactions exist
\cite{D'Hoker:1999ea}. These boundary couplings are in fact responsible 
for extremal n-point functions, namely correlators involving
operators of dimensions $\{ k_1{=}k_2{+}\cdots{+}k_n, k_2, \ldots, k_n\}$.
One must take into account these additional boundary terms
when working out the holographic 1-point functions.

Combining the non-linear, gauge invariant KK reduction map 
with the holographic 1-point functions
we finally arrive at a well defined map between 
the asymptotics of a 10d solution and vevs of gauge invariant 
operators.  

This paper is organized as follows. 
In the next section we 
discuss the Coulomb branch of ${\cal N}=4$ SYM.
We focus on a specific case where the vevs are uniformly
distributed on a disc and compute all vevs of gauge invariant
operators. The challenge for the gravity/gauge theory duality is
to reproduce {\it exactly} these vevs holographically.
In sections 3, 4 and 5 we build the holographic map. In section 3 
we construct gauge invariant variables; in section 4 we work out the 
KK map to second order in the fields and in section 5 we derive the 
holographic 1-point functions. In section 6 we discuss the supergravity
solution dual to the CB state discussed in section 2 and use the 
map developed in sections 3, 4, 5 in order to compute the first 
two non-trivial vevs and find perfect agreement with field theory! 
We conclude in section 7 with a discussion of our results.
Several technical  details are relegated to appendices A, B and C. 
In appendix A
we discuss the harmonic expansion of the antisymmetric gauge field; 
in appendix B we summarize and develop several
results about spherical harmonics with $SO(4)$ symmetry and in 
appendix C we discuss the computation of the field equations 
to second order in fluctuations.

\section{${\cal N}=4$ SYM on the Coulomb branch}

${\cal N}=4$ 
SYM contains 6 scalar fields $X^{i_1}$ in the adjoint representation
of the gauge group that we take to be
$SU(N)$. The Coulomb branch of ${\cal N}=4$ SYM corresponds to giving 
a vacuum expectation value (vev) to the scalars subject to the condition 
$[X^{i_1}, X^{i_2}]=0.$
Upon diagonalizing the scalar fields the moduli space is parametrized
by the $6 (N-1)$ eigenvalues of vevs (modulo the Weyl group). 
In the large $N$ limit we
can approximate the eigenvalues by a
continuous distribution. Notice that the Coulomb branch
still preserves ${\cal N}=4$ supersymmetry but the conformal symmetry and 
the 16 superconformal supersymmetries are broken.
This implies that the vevs are protected from acquiring 
quantum corrections, as we explain at the end of this section.
So this example represents an ideal case 
for a precision test of the gravity/gauge theory correspondence in a 
non-conformal setting. It has been known since \cite{Kraus:1998hv}
that there is a one to one correspondence between the Coulomb branch 
of ${\cal N}=4$ SYM and multicenter D3 brane solutions. Since the vevs
do not renormalize, however, one should be able to establish a 
strong result, namely the exact
values of vevs should be reproducible by a gravitational computation. 

We will consider in this paper the specific case of
a uniform distribution of eigenvalues of $X^1$ and $X^2$
on a disc of radius $a$ and vanishing vev for the 
remaining scalars, $\<X^3\> = \<X^4\>=\<X^5\>=\<X^6\>=0$.
 Let 
\be
X^1 = \r \cos \f, \qquad X^2 = \r \sin \f 
\ee
To leading order in the large $N$ limit we may represent the 
eigenvalues by a uniform continuous distribution,
\be
\s(\r,\f) = \frac{N}{\pi a^2}.
\ee
Notice that the configuration
corresponding to this continuous distribution
preserves an $SO(4) \times SO(2)$ symmetry of $SO(6)$.

To compare with supergravity we would like to parametrize
the moduli space by vevs of composite operators.
We consider the following chiral primaries (CPOs) of ${\cal N}=4$ SYM
\be \label{CPO}
\cao^{I_1} = {\cal N}_{I_1} C^{I_1}_{i_1 \cdots i_k} {\rm{Tr}}(X^{i_1}
\cdots X^{i_k}),
\ee
where ${\cal N}_{I_1}$ is a normalization factor and $C^{I_1}$ is
a totally symmetric traceless rank $k$ tensor of $SO(6)$ which is
normalized such that $\left < C^{I_1} C^{I_2} \right > = C^{I_1}_{i_1
  \cdots i_k} C^{I_2}_{i_1 \cdots i_k} = \d^{I_1 I_2}$. The SYM action
is normalized such that the relevant propagators are 
\be
\left < X^{i}_a(x) X^j_b(y) \right >  = \frac{g_{YM}^2 \d_{ab}
  \d^{ij}}{(2 \pi)^2 \left | x - y \right |^2},
\ee
where $a,b$ are color indices. 

The cases of interest are the operators which are singlets under the
decomposition of $SO(6)$ into $SO(2) \times SO(4)$ since non-singlet
operators have zero vev. These operators can be obtained from the 
explicit expression of scalar harmonics in appendix \ref{harm_ap}
by suitably replacing $x^{i_1}$ by $X^{i_1}$, 
compare  (\ref{CPO}) and (\ref{harm_C}). The result for the 
singlets is
\be
\cao^{2n} = {\cal N}_{2n} \frac{ (-)^{n} }{2^{n} \sqrt{2n+1}}
    {\rm{Tr}} \left ( \sum_{m=0}^{n} (-)^m \left ( \begin{array}{c} n
      \\ m \end{array} \right ) \left ( \begin{array}{c} n + m + 1 
      \\ n + 1 \end{array} \right )  \r^{2m} R^{2(n-m)} \right ).
\ee 
where $R^2 = \sum_{i=1}^6 (X^i)^2$.
The explicit expressions for the lowest dimension operators are thus
\bea
\cao^{2} &=& {\cal N}_{2} \frac{1}{2 \sqrt{3}} 
{\rm{Tr}}(3 \r^2 - R^2); \\
\cao^4 &=& {\cal N}_{4} \frac{1}{4 \sqrt{5}} 
{\rm{Tr}}(10 \r^4 - 8 \r^2 R^2 + R^4); \\
\cao^6 &=& {\cal N}_{6}
\frac{1}{8\sqrt{7}} 
{\rm{Tr}}(35 \r^6 - 45 \r^4 R^2 +15 \r^2 R^4 - R^6).
\eea

To compute the vev of these operators we now use 
\be
\left < {\rm{Tr}}(\r^p) \right > 
= \int_{0}^{a} d\r \r \int_{0}^{2\pi} d\phi 
\s(\r,\phi) \r^p = \frac{2N}{(p+2)} a^p,
\ee
and the identity
\be
(-)^n \left ( \sum_{m=0}^{n} (-)^m \left ( \begin{array}{c} n
      \\ m \end{array} \right ) \left ( \begin{array}{c} n + m + 1 
      \\ n + 1 \end{array} \right )  \right ) = (n+1).
\ee
to arrive at 
\be \la{vev1}
\left < \cao^{2n} \right > = 
\frac{ {\cal N}_{2n} a^{2n}}{ 2^n  \sqrt{2n+1}} N.
\ee

This result was derived by a tree-level computation. However it 
remains uncorrected both perturbatively and non-perturbatively.
A quantum correction to the vev of the scalars $X^i$
would result from a non-vanishing tadpole contribution
and this would induce a correction to the effective potential.
However, there are no perturbative or non-perburtative
quantum corrections to the low energy (2-derivative) effective action 
of $N=4$ SYM \cite{Seiberg,SW} so the vevs of the scalars
are not corrected. The only remaining issue
is operator mixing. Indeed, chiral primary operators mix 
with certain multi-trace operators. However, this is a subleading 
effect\footnote{The only exception is the case of extremal 
operators where the mixing with multitrace operators is not
subleading \cite{D'Hoker:1999ea}. In this 
paper it was argued that the supergravity fields are dual to 
the single trace operators so these are the relevant operators
to consider.} in $1/N$ and we are considering the leading behavior.
It follows that the operators have the same vev (\ref{vev1})
at strong coupling. 
The challenge for the AdS/CFT correspondence is to reproduce these vevs.

\section{KK reduction with gauge invariant variables}

The IIB SUGRA field equations\footnote{The field strength differs by a 
factor of 4 from the conventions in \cite{Polchinski:1998rr}. 
Index conventions:
$M,N,...$ are $10d$ indices, $\m,\n,...$ are $AdS_5$ indices,
$a,b,...$ are $S^5$ indices. $x$ denotes AdS coordinates and $y$ 
$S^5$ coordinates.} for the metric and 5-form field strength are given by:
\be 
R_{MN} = \frac{1}{6} F_{MPQRS} F_M{}^{PQRS}, \qquad
F=*F.
\ee
These equations admit an $AdS_5 \times S^5$ solution
\bea
ds_o^2 &=& \frac{dz^2}{z^2} + \frac{1}{z^2} d x_{||}^2 
+ d\q^2 + \sin^2 \q d \W_3^2 + \cos^2 \q d \f^2 \\
&&F^{o}_{\m \n \r \s \t} = \e_{\m \n \r \s \t}, \qquad
F^{o}_{abcde} = \e_{abcde} \nn
\eea
We will consider here solutions that are deformations of
$AdS_5 \times S^5$ such that 
\bea
g_{MN} &=& g^o_{MN} + h_{MN}, \\
F_{MNPQR} &=&  F^o_{MNPQR} + f_{MNPQR}.\nn
\eea
The fluctuations can be expanded in $S^5$ harmonics:
\bea \label{fluct_h}
h_{\m \n}(x,y) &=& \sum \tilde{h}^{I_1}_{\m \n}(x) Y^{I_1}(y) \nonumber \\
h_{\m a} (x,y)&=& 
\sum (\tilde{B}^{I_5}_{(v)\m}(x) Y_a^{I_5}(y) 
+ \tilde{B}^{I_1}_{(s)\mu}(x) D_a Y^{I_1}(y)) \nonumber \\
h_{(ab)}(x,y) 
&=& \sum (\hat{\phi}_{(t)}^{I_{14}}(x) Y_{(ab)}^{I_{14}}(y) 
+ \phi^{I_5}_{(v)}(x) D_{(a} Y^{I_5}_{b)}(y)
+ \phi^{I_1}_{(s)}(x) D_{(a} D_{b)} Y^{I_1}(y) )\nonumber \\
h_{a}^a(x,y) &=& \sum \tilde{\pi}^{I_1}(x) Y^{I_1}(y) 
\eea
and 
\bea
f_{\m \n \r \s \t}(x,y) &=& \sum
5 D_{[\m} b^{I_1}_{\n \r \s \t]}(x) Y^{I_1}(y) \nonumber \\
f_{a \m \n \r \s}(x,y) &=& \sum (b^{I_1}_{\m \n \r \s}(x) D_a Y^{I_1}(y) 
+ 4 D_{[\m} b^{I_5}_{\n \r \s]}(x) Y_a^{I_5}(y)) \nonumber \\
f_{ab \m \n \r}(x,y) &=& \sum
(3 D_{[\m} b^{I_{10}}_{\n \r]}(x) Y_{[ab]}^{I_{10}}(y) 
-2 b^{I_5}_{\m \n \r}(x) D_{[a} Y_{b]}^{I_5}(y)) \nonumber \\
f_{a b c \m \n}(x,y) &=& \sum 
(2 D_{[\m} b_{\n]}^{I_5}(x) \e_{abc}{}^{de} D_d Y_e^{I_5}(y)
+ 3 b_{\m \n}^{I_{10}}(x) D_{[a} Y_{bc]}^{I_{10}}(y)) \nonumber \\
f_{abcd\m}(x,y)  &=& \sum 
(D_\m b_{(s)}^{I_1}(x) \e_{abcd}{}^e D_e Y^{I_1}(y)
+ (\L^{I_5}-4) b_\m^{I_5}(x) \e_{abcd}{}^e Y_e^{I_5}(y)) \nn \\
f_{a b c d e}(x,y) &=& \sum b_{(s)}^{I_1}(x) \L^{I_1} \e_{abcde} Y^{I_1}(y) 
\eea
Numerical constants in these expressions are inserted so 
as to match with the conventions of \cite{Kim:1985ez},
see appendix \ref{5form}.
Parentheses denote a symmetric traceless combination
(i.e. $A_{(ab)} = 1/2 (A_{ab}+A_{ba}) -1/5 g_{ab} A_a^a$).
$Y^{I_1}, Y_a^{I_5}, Y_{(ab)}^{I_{14}}$ and $Y_{[ab]}^{I_{10}}$ denote scalar, 
vector and tensor harmonics whilst $\L^{I_1}$ and $\L^{I_5}$ are the 
eigenvalues of the scalar and vector harmonics under (minus) the d'Alembertian.
The subscripts $t$, $v$ and $s$ denote whether the field is associated with
tensor, vector or scalar harmonics respectively, whilst the
superscript of the harmonic label $I_n$ derives from the number of
components $n$ of the harmonic.

Not all fluctuations are independent however. Some of the 
modes are diffeomorphic to each other or to the background solution,
i.e. certain $\delta h_{MN}$ and $\d f_{MNPQR}$ are generated by 
a coordinate transformation,
\be
x^M{}' = x^M - \xi^M.
\ee
These, up to terms linear in fluctuations, are given by
\bea \label{gau_tr}
\d h_{MN} &=& (D_M \xi_N +  D_N \xi_M) + (D_M \xi^P h_{PN} + D_N \xi^P h_{MP}
+\xi^P D_P h_{MN}); \label{diffeo} \\
\d f_{MNPQR} &=& 5 D_{[M} \xi^S F^o_{NPQR]S} +  (5 D_{[M} \xi^S f_{NPQR]S}
+\xi^S D_S f_{MNPQR}). \nonumber
\eea
The gauge parameter $\xi^M(x,y)$ can be expanded in harmonics as
\bea
\x_\mu(x,y)&=&\sum \xi^{I_1}_\mu(x) Y^{I_1}(y); \\
\x_a(x,y) &=&\sum (\xi_{(v)}^{I_5}(x) Y_a^{I_5}(y) 
+ \xi_{(s)}^{I_1}(x) D_a Y^{I_1}(y)). \nonumber
\eea
In much of the previous literature this issue was dealt with by 
imposing a gauge fixing condition, most notably the 
de Donder-Lorentz gauge fixing condition
\be \label{donder}
D^a h_{(ab)}=D^a h_{a\m}=0.
\ee
This amounts to setting to zero the coefficients
$\tilde{B}_{(s)\mu}^{I_1}, \f_{(v)}^{I_5}, \f_{(s)}^{I_1}$
(as can be easily seen by inserting (\ref{fluct_h}) in (\ref{donder})). 
Although this gauge is a very convenient choice for deriving the 
spectrum, it is not very well suited for holography since generically
solutions will not be in that gauge. For this reason instead of 
gauge fixing this symmetry we will derive gauge invariant 
combinations of fluctuations. This will allow us to switch easily
between different gauges. 

\subsection{Gauge invariance at linear order}

We first discuss gauge invariance at leading order, i.e.
we consider the fluctuation independent terms in (\ref{diffeo}).
These transformations map the fluctuations to the background
solution. Under these transformations the coefficients 
in (\ref{fluct_h}) transform as
\bea \label{lin_tr}
&& \d \tilde{h}_{\m \n}^{I_1} = D_\m \xi_\n^{I_1} +  D_\n \xi_\m^{I_1}, \qquad
\d \tilde{B}_{(v)\m}^{I_5} = D_\mu \xi_{(v)}^{I_5}, \qquad 
\d \tilde{B}_{(s)\m}^{I_1} = D_\mu \xi_{(s)}^{I_1} + \xi_\m^{I_1}, \nn \\ 
&& \qquad \d \hat{\f}_{(t)}^{I_{14}}=0, \qquad
\d \f_{(v)}^{I_5} = 2 \xi_{(v)}^{I_5},
\qquad \d \f_{(s)}^{I_1} = 2 \xi_{(s)}^{I_1}, \qquad 
\d \tilde{\pi}^{I_1} = 2 \L^{I_1} \xi_{(s)}^{I_1}.
\eea
It follows that $\hat{\f}_{(t)}^{I_{14}}$ is gauge invariant to this 
order and for the rest of fields
we can construct the following gauge invariant combinations
\bea \label{inv_lin}
\hat{\pi}^{I_1} &=& \tilde{\pi}^{I_1} - \L^{I_1} \f_{(s)}^{I_1} \\
\hat{B}^{I_5}_{(v) \mu} & =& \tilde{B}^{I_5}_{(v)\mu} 
- \frac{1}{2} D_\mu \f^{I_5}_{(v)}  \nn \\
\hat{h}^{I_1}_{\m \n} &=& \tilde{h}_{\m \n}^{I_1} 
- D_{\m} \hat{B}^{I_1}_{(s)\n}
-D_\n \hat{B}^{I_1}_{(s)\m}, 
\qquad I_1 \neq 0. \nn
\eea
where we define
\be \label{bhat_def}
\hat{B}_{(s)\mu}^{I_1} = \tilde{B}^{I_1}_{(s)\mu} 
- \frac{1}{2} D_\mu \f^{I_1}_{(s)} \qquad \Rightarrow \qquad
\d \hat{B}_{(s)\mu}^{I_1} = \xi_\m^{I_1}.
\ee
Note that the last formula in (\ref{inv_lin})
is only valid for $I_1{\neq}0$ 
(the fields $\tilde{B}_{(s)\mu}^{I_1}$ and $\phi_{(s)}^{I_1}$
exists only for $I_1>0$, since $Y^0$ is a constant). For $I_1=0$,
$\tilde{h}^0_{\m \n}$ is a deformation of the background metric 
and from (\ref{lin_tr}) we see that it indeed transforms as a metric.

Similarly, the leading term in the 5-form transformation implies for the 
coefficients in the harmonic expansion the following transformations
\bea
&&\d b^{I_1}_{(s)} = \xi_{(s)}^{I_1}, \qquad 
\d b_{\m \n \r \s}^{I_1} = \e_{\m \n \r \s}{}^\t \xi_\t^{I_1}, \qquad 
\d b_\m^{I_5} = \frac{1}{(\L^{I_5}-4)} D_\m \xi_{(v)}^{I_5}, \nn \\
&& \d b_{\m \n \r}^{I_5} = \d b_{\m \n}^{I_{10}}=0,
\eea
so that the gauge invariant combinations are 
\bea
\hat{b}^{I_1} &=& b^{I_1}_{(s)} - \frac{1}{2} \f_{(s)}^{I_1} \\
\hat{b}{}^{I_1}_{\m \n \r \s} &=& b_{\m \n \r \s}^{I_1}
- \e_{\m \n \r \s}{}^\t \hat{B}_{(s)\t}^{I_1} \nn \\
\hat{b}{}^{I_5}_{(v)\m} &=& b_{(v)\m}^{I_5} - \frac{1}{2 (\L^{I_5}-4)} D_\m 
\f_{(v)}^{I_5}.
\eea
and the fields $b_{\m \n \r}^{I_5}$ and $b_{\m \n}^{I_{10}}$.

\subsection{Gauge invariance at quadratic order}

In this subsection we will derive 
the gauge invariant combinations to second
order in the fluctuations. The idea is the same as in the previous 
subsection: we insert the harmonic expansion of the fluctuations 
and the gauge parameter into (\ref{gau_tr}) (which now includes
all terms) and read off the transformation of each coefficient. 
Then we seek a quadratic modification of each field combination
that is gauge invariant. One complication in this case is that 
because the r.h.s. of (\ref{gau_tr}) is non-linear one needs to 
project onto the basic of spherical 
harmonics in order to extract the transformation
of the coefficients. The analysis can be readily carried out in all 
generality but for the applications considered in this 
paper it is sufficient to consider only 
the modes that couple to scalar harmonics and their derivatives,
i.e. we set to zero all fields that couple to vector 
and tensor harmonics (and their derivatives) (as well as 
$\xi_{(v)}^{I_5}$). Including these fields
would result in additional terms in the gauge invariant combinations below.
Since no confusion can arise we also drop the subscript $(s)$ from relevant 
fields and use the condensed notation for indices and 
arguments: 
$\f_{(s)}^{I_1} \to \f^1, \tilde{\pi}^{I_1} \to \tilde{\pi}^1, 
z(k_1) \to z_1, a(k_1,k_2,k_3) \to a_{123}$ etc. Note that we
consistently use the notation $\td{\psi}$ to denote a field in the
harmonic expansion of the supergravity field; $\hat{\psi}$ to denote a
field which is gauge invariant to linear order and $\psi$ to denote
the field which is gauge invariant to quadratic order.
 
\subsubsection{Scalar fields}

We first discuss the scalar fields, $\tilde{\pi}^{I_1}, \f_{(s)}^{I_1}$ and
$b_{(s)}^{I_1}$.
Their transformations are given by (we suppress the linear terms
determined in the previous section and factors
$\< \cC^{I_1}\cC^{I_2}\cC^{I_3} \>$):
\bea
\d \tilde{\pi}^{1} &=&\frac{1}{z_1} 
\left( 2 \f^{2} \xi^{3}  d_{123}
+(\frac{2}{5} \L^{2} \xi^{2} \tilde{\pi}^{3} 
+ \xi^{\mu 2} D_\mu \tilde{\pi}^{3}) a_{123}
+ (\xi^{2} \tilde{\pi}^{3} 
+ 2 \xi^{\mu 2} \tilde{B}_{\mu}^{3} )
b_{123} \right); \\
\d \f^{1} &=& \frac{1}{z_1 q_1}
\left(\xi^{2} \f^{3} e_{123}
+ (\xi^{\mu 2} D_\mu \f^{3} 
+ \frac{2}{5} \xi^{2} \tilde{\pi}^{3}) d_{213}
+ 2 \xi^{\mu 2} \tilde{B}_{\mu}^{3} c_{123}
\right); \\
\d b^{1} &=& \frac{1}{\L^{1} z_1} \left(
(\xi^{\mu 2} D_\mu b^{3} + \L^{2}  b^{2} \xi^{3} )
(b_{123} + \L^{3} a_{123})\right),
\eea
where the triple overlaps $a_{123}=a(k_1,k_2,k_3), b_{123}=b(k_1,k_2,k_3)$ etc 
are defined in appendix 
\ref{harm_ap} and summation over $(I_2,I_3)$ is implicit.

From these transformations one can infer quantities which are gauge invariant
to quadratic order:
\bea
\pi^{1} &=& \hat{\pi}^{1} - \frac{1}{2 z_1}\left( 
\left( \frac{2}{5} \Lambda^{2} a_{123} + b_{123}
- \frac{2 \Lambda^{1} }{5 q_1} d_{213} \right) \f^{2} \hat{\pi}^{3} 
+ \left(d_{123} -\frac{\Lambda^{1}}{2 q_1} e_{123} + 
\right . \right. \la{pi1}   \\
&& \left . \left.
\Lambda^{3}  (\frac{1}{5} \Lambda^{2} a_{123} + \frac{1}{2} b_{123}
- \frac{\Lambda^{1}}{5 q_1} d_{213} ) \right ) \f^{2} \f^{3} 
+ 2 \hat{B}_{\mu}^{2}
\left (D^{\mu} \hat{\pi}^{3}a_{123} + \hat{B}^{3\mu} (b_{123} - 
\frac{2 \Lambda^{1}}{q_1} c_{123}) \right) \right); \nn \\
b^{1} &=& \hat{b}^{1} +  \frac{1}{z_1} \left(
\frac{\Lambda^{3}}{2 \Lambda^1} 
\f^{2} \hat{b}^{3} b_{312}
+ \frac{1}{10 q_1} d_{213} \f^{2} \hat{\pi}^{3}
+  \left (\frac{\Lambda^{3}}{8 \Lambda^{1}} b_{312}+ 
\frac{\Lambda^{1}}{20 q_1} d_{213} + \frac{1}{8 q_1} 
e_{123} \right ) \phi^{2} \phi^{3} \right.
\nn \\
&& \left.  + \hat{B}_{\mu}^{2} 
\left ( \frac{1}{2 q_1} \hat{B}^{3\mu} c_{123}
+ \frac{1}{\Lambda^{1}} D^{\mu} \hat{b}^{3}  b_{213} \right) 
\right ). \la{b1} 
\eea
The $I_1=0$ sector is special because the scalar harmonic is 
constant. Notice that we only have one scalar in this sector,
namely $\tilde{\pi}^0$. Working out the gauge transformation 
yields
\be
\d \tilde{\pi}^0 = z(k) \left(2 \xi^{I} \f^{I} q(k)
+\frac{2}{5} \L^{I} \xi^{I} \tilde{\pi}^I 
+ \xi^{\mu I}  D_\mu \pi^I 
-(\xi^I \tilde{\pi}^I + 2 \xi^{\mu I} B_{\mu}^I) \L^I \right).
\ee
From here we obtain that the gauge invariant combination is
(notice that $\tilde{\pi}^0$ was gauge invariant to leading order, 
i.e. $\hat{\pi}^0 = \tilde{\pi}^0$) 
\be \label{pi0}
\pi^0 = \tilde{\pi}^0 + z(k)\left(\frac{3}{10} \L^I  \f^I \hat{\pi}^I
-\frac{1}{4} \L^I(\L^I+8) \f^I \f^I 
- \hat{B}^{\mu I} D_{\mu} \hat{\pi}^I 
+ \L^I \hat{B}^{\mu I} \hat{B}_{\mu}^{I}
\right),
\ee
where the summation over $I$ is implied and 
$\hat{\pi}^I$ and $\hat{B}^{\mu I}$ are defined in 
(\ref{inv_lin}) and (\ref{bhat_def}). 

Let us now consider the field $\hat{\f}_{(t)}$ associated with the tensor
harmonic. Whilst this is gauge invariant to leading order, at the next
order it transforms as
\be
\d \hat{\f}^{1}_{(t)} = \frac{1}{z_{(t)1}} \left ( \xi^2 \f^3 e^{(t)}_{123} 
- (\xi^{\m 2} D_{\m} \f^3 + \frac{2}{5} \xi^2 \td{\pi}^3 -
2 \xi^{\m2} \td{B}_{\m}^3) c^{(t)}_{123} \right),
\ee
where the normalization factor 
$z_{(t)}$ and overlap integrals $c^{(t)}_{123}$ etc are defined in
the appendix \ref{harm_ap}. Thus the gauge invariant combination to
this order is
\be \label{phit}
\f^1_{(t)} = \hat{\f}^1_{(t)} + \frac{1}{z_{(t)1}}
\left ( (-\hat{B}^{\m2} \hat{B}_{\m}^3 + \frac{1}{5} \hat{\pi}^2 \f^3 
+ \frac{1}{10} \L^3 \f^2 \f^3) c^{(t)}_{123} - \frac{1}{4} 
e^{(t)}_{123} \f^2 \f^3 \right ).
\ee

\subsubsection{Tensor fields} \label{g2ndt}

We now turn to the non-scalar sector. As we will see in the next section
the field equations algebraically relate the field $b_{\mu \n \r \s}^{I_1}$
to the field $b_{(s)}^{I_1}$ (more precisely the field
equations relate the corresponding gauge
invariant combinations) so we need not discuss this field. Furthermore, 
$\tilde{B}_{(s)\mu}^{I_1}$ is pure gauge. It is useful however to 
introduce the following combination that transforms nicely up to 
quadratic order
\bea
B_{\mu}^1 &=& \hat{B}_{\m}^1 
+ \frac{1}{z_1}\left(
- \frac{1}{2} (\frac{1}{10} D_\mu \f^2 \hat{\pi}^3 
+ \hat{B}^{\nu 2} \hat{h}_{\m \n}^3) a_{123}
\right. \\
&&\hspace{-1cm}\left. +D_\mu \left( \f^2 \f^3 
(- \frac{1}{8} b_{123} +\frac{1}{8 q_1} e_{123} +\frac{\L^3}{5 q_1} d_{213})
+\frac{1}{10 q_1} d_{213} \f^2 \hat{\pi}^3 
+ \frac{1}{2} (\frac{1}{q_1} c_{123} - a_{123}) 
\hat{B}_{\n}^2 \hat{B}^{\nu 3}\right)
\right).\nn
\eea
This transforms as
\be
\d B_{\mu}^1 = \xi_\mu^1 
+ \frac{1}{z_1}(\xi^2 \hat{B}_{\m}^3 b_{123} 
+ \xi^{\n 2} D_\n \hat{B}_{ \m}^3 a_{123}).
\ee 
Now consider the KK graviton modes, 
$\tilde{h}_{\m \n}^{I_1}$. The gauge transformation reads
\be
\d \tilde{h}_{\m \n}^{1} = \frac{1}{z_1}\left(
(D_\mu \xi^{\l 2} \tilde{h}_{\l \n}^{3}+
D_\nu \xi^{\l 2} \tilde{h}_{\l \m}^{3}+
\xi^{\l 2} D_\l \tilde{h}_{\m \n}^{3}) a_{123} 
+(\xi^{2} \tilde{h}_{\m \n}^{3} 
+ 2 D_{(\mu} \xi^{2} \tilde{B}^3_{\nu)}) b_{123}\right).
\ee
From this we derive the following gauge 
invariant combination ($I_1 \neq 0$)
\bea
h_{\m \n}^1 &=& \tilde{h}_{\m \n}^1 - D_{\m} B_{\n}^1 - D_{\n} B_{\m}^1 
-\frac{1}{z_1} \left(\frac{1}{2}(\f^2 \hat{h}_{\m \n}^3 
+ \frac{1}{2} D_\m \f^2 D_\n \f^3) b_{123} \right. \\
&+&\left.(D_{\m} \hat{B}^{\l 2} \hat{h}_{\n \l}^3 + 
D_{\n} \hat{B}^{\l 2} \hat{h}_{\m \l}^3+
\hat{B}^{\l 2} D_\l \hat{h}_{\m \n}^3 
+ D_\m \hat{B}^{\l 2} D_\n \hat{B}_{\l}^3 + 
\hat{B}^{\l 2} \hat{B}_{\l}^3 g_{\m \n}^0
-\hat{B}_{\m}^2 \hat{B}_{\n}^3) a_{123} 
\right). \nn
\eea
Let us now discuss the $I_1=0$ case. $\tilde{h}_{\m \n}^0$ transforms as
\bea
\d \tilde{h}_{\m \n}^{0} &=& D_\mu \xi^{\l 0} \tilde{h}_{\l \n}^{0}+
D_\nu \xi^{\l 0} \tilde{h}_{\l \m}^{0}+
\xi^{\l 0} D_\l \tilde{h}_{\m \n}^{0} \\ 
&+& z(k)\left(
D_\mu \xi^{\l I} \tilde{h}_{\l \n}^{I}+
D_\nu \xi^{\l I} \tilde{h}_{\l \m}^{I}+
\xi^{\l I} D_\l \tilde{h}_{\m \n}^{I} 
- \L^I (\xi^{I} \tilde{h}_{\m \n}^{I} 
+ 2 D_{(\mu} \xi^{I} \tilde{B}_{\nu)}^I) \right). \nn
\eea
We introduce
\bea \label{ginv_h0}
h_{\m \n}^0 &=& \tilde{h}_{\m \n}^0 +\frac{1}{3} \pi^0 g_{\m \n}^o
- z(k) \left(\frac{1}{2}\L^I (\f^I \hat{h}_{\m \n}^I 
+ \frac{1}{2} D_\m \f^I D_\n \f^I)  \right. \\
&&\left. + D_{\m} \hat{B}^{\l I} \hat{h}_{\n \l}^I + D_{\n}
\hat{B}^{\l I} \hat{h}^{I}_{\m \l}  +
\hat{B}^{\l I} D_\l \hat{h}_{\m \n}^I 
+ D_\m \hat{B}^{\l I} D_\n \hat{B}_{\l}^I + 
\hat{B}^{\l I} \hat{B}_{\l}^I g_{\m \n}^o
-\hat{B}_{\m}^I \hat{B}_{\n}^I \nn
\right)
\eea
(the term linear in $\pi^0$ was added in anticipation of the 
fact that it is $\tilde{h}_{\m \n}^0 +\frac{1}{3} \tilde{\pi}^0 g_{\m \n}^o$
that satisfies the linearized equations of motion, see the discussion 
around (\ref{grav_lin})).
Recall that this mode is a correction to the spacetime metric
\be \label{corr_met}
g_{\m \n} = g^o_{\m \n} + h_{\m \n}^0,
\ee
so the combination should transform as 
\be
\d g_{\m \n} = D^{g}_{\m} \z_{\n} + D^{g}_{\n} \z_{\m},
\ee
where $D^{g}$ is the covariant derivative of the corrected
metric (\ref{corr_met}).
Indeed one finds this to hold with 
\be
\z_{\n} = \xi^{\m 0} g_{\m \n} +
z(I) (\xi^{\l I} D_\l \hat{B}_\n^I - \L^I \xi^I \hat{B}_\n^I).
\ee

\section{Field equations}

The field equations for the coefficients in the harmonic expansion
were derived in the de Donder gauge at linear order by
\cite{Kim:1985ez} and at quadratic order 
in \cite{LMRS,Arutyunov:1999en,Lee:1999pj}.
The gauge invariant variables derived in the previous section 
allow one to relax the gauge condition. Indeed notice that the 
gauge invariant variables when evaluated in the de Donder gauge 
become equal to the fields used in  
\cite{Kim:1985ez,LMRS,Arutyunov:1999en,Lee:1999pj}.
It follows (and we have explicitly checked this in detail) that the field 
equations with no gauge condition imposed can be obtained
from the results in \cite{Kim:1985ez,LMRS,Arutyunov:1999en,Lee:1999pj}
by simply replacing each field with its gauge invariant generalization.

\subsection{Linear order} \label{lin_ord}

In this subsection we summarize some of the results of \cite{Kim:1985ez}
(a summary of the derivation is given in appendix
\ref{2ndorder}).
As just mentioned, one can relax the gauge fixing condition by 
replacing all fields by the hatted versions given in the previous
section. 

The scalars satisfy the following equations
\bea
\Box \hat{s}^{I_1} &=& k (k-4) \hat{s}^{I_1},  \quad k \geq 2, \nn \\
\Box \hat{t}^{I_1} &=& (k+4) (k+8)\hat{t}^{I_1}, \quad k \geq 0, \\
\Box \hat{\f}^{I_{14}}_{(t)} &=& k (k+4) \hat{\f}_{(t)}^{I_{14}} 
\quad k \geq 2, \nn
\eea
where we introduce the combinations
\be \la{l1}
\hat{s}^{I_1} = \frac{1}{20 (k+2)} (\hat{\pi}^{I_1} 
- 10 (k+4) \hat{b}^{I_1}), \qquad
\hat{t}^{I_1} = \frac{1}{20 (k+2)} (\hat{\pi}^{I_1} + 10 k \hat{b}^{I_1}),
\ee
with inverse relations
$\hat{b}^{I_1}=-\hat{s}^{I_1}+\hat{t}^{I_1},\ 
\hat{\pi}^{I_1}=10k\hat{s}^{I_1}+10(k+4)\hat{t}^{I_1}$.

The remaining modes that couple to scalar spherical harmonics are
the KK gravitons. They are described by transverse and traceless
fields
\be
\f_{(\m \n)}^{I_1} = \hat{h}^{I_1}_{(\m \n)} - \frac{1}{(k+1)(k+3)} 
D_{(\m} D_{\n)} (\frac{2}{5} \hat{\pi}^{I_1} - 12 \hat{b}^{I_1}), 
\quad I_1 \neq 0.
\ee
satisfying the equation,
\be
(\Box -(k(k+4)-2)) \f_{(\m \n)}^{I_1} =0, \quad I_1 \neq 0.
\ee
The $I_1=0$ case is special in that this mode describes a deformation
of the background metric. The combination that satisfies the $5d$ linearized
Einstein equation is
\be \label{grav_lin}
h_{\m \n}'{}^0 = (\tilde{h}_{\m \n}^0 + \frac{1}{3} g_{\m \n}^o \tilde{\pi}^0).
\ee
One can understand the origin of the shift by $\tilde{\pi}^0$ by considering
the reduction of the $10d$ action to five dimensions. Keeping terms 
linear in the fluctuations, the volume of compactification manifold is
\be
\int d^5 y \sqrt{\det g_{ab}} = \pi^3 (1 + \frac{1}{2} \tilde{\pi}^0)
\ee
It follows that the reduced action is 
\be
S_{5d} \sim \int d^5 x \sqrt{\det g_{\m \n}} ((1 + 
\frac{1}{2} \tilde{\pi}^0) R + \cdots)
\ee
and a Weyl transformation is required to bring the action to the 
Einstein frame. The transformation from $\tilde{h}_{\m \n}^0$ to 
$h_{\m \n}'{}^0$ is precisely this Weyl transformation.

\subsection{Quadratic order}

The derivation of the equations of motion to second order in fluctuations 
was discussed in \cite{LMRS,Arutyunov:1999en,Lee:1999pj} and is 
summarized in appendix \ref{2ndorder}. For our applications it is 
sufficient to retain only the quadratic coupling to
the field $s^2$.

\subsubsection{Scalar fields}

The corrected field equation for the scalar fields 
$\psi=\{s^2, s^4, t^0, t^2, t^4,\f_{(t)}^2\}$ is given by
\be \label{sc_cor}
(\Box - m^2_{\psi}) \psi^I = D_{\psi22} (\hat{s}^2)^2 
+ E_{\psi22} D_{\m} \hat{s}^2 D^{\m} \hat{s}^2
+ F_{\psi22} D_{(\m} D_{\n)} \hat{s}^2 D^{(\m} D^{\n)} \hat{s}^2,
\ee
where the coefficients $D_{\psi22}, E_{\psi22}, F_{\psi22}$ can be 
obtained from the results in appendix \ref{2ndorder} and are given in
table 1.
\begin{table}
\begin{center}
\begin{tabular}{|c|c|c|c|c|c|c|}
\hline
\  & $s^2$ & $s^4$ & $t^0$ & $t^2$ & $t^4$ & $\f_{(t)}^2$ \\
\hline \hline
$D_{\psi 22}$ & $- \frac{4 \sqrt{3}}{3}$ & $ -\frac{172}{5 \sqrt{5}}$
& $\frac{229}{75}$ &
$\frac{76 \sqrt{3}}{25}$ & $\frac{52}{ \sqrt{5}}$ & $\frac{48}{25}$\\
$E_{\psi 22}$ & $\frac{\sqrt{3}}{10}$ & $\frac{3}{\sqrt{5}}$ & $-\frac{11}{20}$ &
$-\frac{3 \sqrt{3}}{10}$ & $-\frac{1}{\sqrt{5}}$ & $-\frac{4}{5}$\\
$F_{\psi 22}$ & $\frac{1}{12 \sqrt{3}}$ & $\frac{7}{9 \sqrt{5}}$  & $\frac{1}{60}$ &
$\frac{\sqrt{3}}{180}$ & $0$ & $\frac{2}{45}$\\
\hline
$A_{\psi s \f}$ & $\frac{7 \sqrt{3}}{40}$ & $\frac{7}{2 \sqrt{5}}$ 
& $-\frac{3}{40} $
& $-\frac{7 \sqrt{3}}{120}$ & $-\frac{1}{2 \sqrt{5}}$ & $-\frac{1}{5}$\\
$A_{\psi \f \f}$ & $-\frac{17 \sqrt{3}}{160}$ & $-\frac{9}{10
  \sqrt{5}}$ 
& $-\frac{1}{80}$ & $\frac{3 \sqrt{3}}{160}$ & $\frac{1}{8 \sqrt{5}}$ &
$\frac{1}{20}$ \\
$A_{\psi s B}$ & $-\frac{\sqrt{3}}{24}$ & $-\frac{3}{4\sqrt{5}}$ 
& $-\frac{1}{48}$ & $-\frac{\sqrt{3}}{120}$ & $0$ & $0$ \\
$A_{\psi B B}$ & $-\frac{\sqrt{3}}{32}$  & $-\frac{1}{4 \sqrt{5}}$ 
 & $-\frac{1}{80}$ & $-\frac{\sqrt{3}}{160}$ & $0$ & $\frac{1}{20}$ \\
\hline
$J_{\psi 22}$ & $\frac{-2 \sqrt{3}}{15}$ & $-\frac{83}{18 \sqrt{5}}$ &
$\frac{3}{40}$ 
& $\frac{2 \sqrt{3}}{45}$ & $\frac{1}{2 \sqrt{5}}$ & $\frac{2}{45}$ \\
$L_{\psi 22}$ & $-\frac{1}{12 \sqrt{12}}$ & $-\frac{7}{18\sqrt{5}}$ &
$-\frac{1}{120}$  &
$-\frac{\sqrt{3}}{360}$  & $0$ & $-\frac{1}{90}$ \\
$w(\psi)$ & $\frac{\sqrt{8}}{3}$ & $\frac{2 \sqrt{3}}{5}$  & $\frac{8
  \sqrt{5}}{\sqrt{3}}$ & $\frac{4 \sqrt{7}}{\sqrt{10}}$  &
$\frac{12}{\sqrt{70}}$ & $\frac{\sqrt{15}}{4}$ \\
\hline
$\l_{\Psi 22}$ &$- \frac{4}{\sqrt{6}}$ & 0 & 0 &0 &0 &  0\\
\hline
\end{tabular}
\caption{Coefficients in (\ref{sc_cor}), (\ref{psi_ginv}), (\ref{10to5})
and (\ref{5deqn}).}
\end{center}
\end{table}
The fields entering the l.h.s of this equation are the gauge invariant 
combinations to second order whilst the fields in the r.h.s. are the 
gauge invariant combinations to linear order (since the r.h.s. is 
quadratic in fluctuations). This follows from our general discussion
and we have also explicitly checked that 
the terms involving $\f_{(s)}^2$ in the second order equations (with no 
gauge fixing imposed) are accounted for by the $\f_{(s)}^2$ terms 
in the gauge invariant combinations. When $s^2$ is the leading
non-zero field 
(as it is in the application discussed in this paper) the 
gauge invariant combinations take the form
\be \label{psi_ginv}
\psi = \hat{\psi} + A_{\psi s\phi} \hat{s}^2 \phi_{(s)}^2 
+ A_{\psi \f \f} (\f_{(s)}^2)^2 + A_{\psi s B} D^\m \hat{s}^2 \hat{B}_{(s)\m}^2
+ A_{\psi B B} \hat{B}_{(s)}^{\m 2} \hat{B}_{(s)\m}^2. 
\ee
The coefficients $ A_{\psi s\phi}, A_{\psi \f \f}, 
A_{\psi s B}$ and
$A_{\psi B B}$ are given in Table 1. 

The field equations in (\ref{sc_cor}) contain higher derivative 
terms on the r.h.s which can however be removed 
by the following transformation \cite{LMRS}
\be \label{10to5}
\Psi = w(\psi) \left(\psi + J_{\psi22} (\hat{s}^2)^2 
+ L_{\psi22} D_\mu \hat{s}^2 D^\mu \hat{s}^2 \right).
\ee
This transformation {\it is} the non-linear KK map to quadratic 
order in the fields. It maps solutions of the 10d fields equations 
to solutions of the 5d field equations,
\be \label{5deqn}
(\Box - m_{\psi}^2)\Psi = \l_{\Psi 22} (S^2)^2.
\ee
The coefficients $w(\psi), J_{\psi22}, L_{\psi22}$ and $l_{\Psi 22}$
are given in Table 1. We include on the r.h.s. only the 
terms quadratic in $S^2$ because these are the terms that are relevant 
for us. We note however that all quadratic terms
(and cubic scalar couplings \cite{Arutyunov:1999en})
have been determined in the literature 
\cite{LMRS,Arutyunov:1999en,Lee:1999pj}. The results in Table 1 are 
in agreement with the results in these papers. 
The field equations can be integrated to a $5d$ action. 
and the constants $w(\psi)$ have been chosen such that the 
overall normalization agrees with the one in \cite{BFS1,BFS2}
\be \label{5daction}
S_{5d} = 
\frac{N^2}{2 \pi^2} \int d^5 x \sqrt{G}(\frac{1}{4} R+
\frac{1}{2} G^{\m \n} \pa_\m \Psi \pa_\n \Psi + V(\Psi)).
\ee
Using the quadratic 10-dimensional supergravity action computed in
\cite{LMRS,Arutyunov:1998hf} one finds that
\bea
w(s^I) &=& \sqrt{ \frac{8 k (k-1) (k+2) z(k)}{(k+1)}}, \qquad
w(\f_{(t)}) = \sqrt{ \frac{z_{(t)}(k^{(t)})}{8}} \\
w(t^I) &=& \sqrt{ \frac{8 (k+2)(k+4)(k+5) z(k)}{(k+3)}}.  \nn
\eea
From Table 1 we see that the only non-zero cubic term in the
potential (that is also quadratic in $S^2$) is the cubic self 
coupling of $S^2$ and its coefficient, $-4/3 \sqrt{6}$, 
precisely agrees with the corresponding coupling in $5d$ gauged 
supergravity (after matching sign conventions), 
compare with (2.6) of \cite{BFS2}.

It is important to note that the transformation (\ref{10to5}) gives 
an explicit map between  
solutions of the ten dimensional equation and solutions of the 
five dimensional equation and vice versa, i.e.
any solution of the five dimensional
theory specified by the action (\ref{5daction}) can be uplifted
to a ten dimensional solution. We emphasize that this map is valid 
irrespectively of whether there is a consistent truncation since we keep all 
KK modes.

\subsubsection{Tensor fields}

Let us consider first the graviton.
The quadratic correction to the gravitational equation 
is obtained in appendix \ref{2ndorder}:
\bea
(L_E + 4) h_{\m \n}^0 &=& \frac{1}{12} \left (
-\frac{2}{9} D_\m D_\r D_\s \hat{s}^2  D_\n D^\r D^\s \hat{s}^2 
-\frac{16}{3} (D_\m D_\n D_\r \hat{s}^2 D^{\r} \hat{s}^2
+D_\m D_\r \hat{s}^2 D_\n D^{\r} \hat{s}^2)  \right . \nn \\
&& \left . \hspace{-1cm}+\frac{364}{9} D_\m \hat{s}^2 D_\n \hat{s}^2
+g_{\m \n}^o\left(-\frac{8}{9} D_\r D_\s \hat{s}^2 D_\r D_\s \hat{s}^2 
+ 20 D_\r \hat{s}^2 D_\r \hat{s}^2 -\frac{512}{9} s^2 \right) \right ),
\eea
where $L_E$ is the linearized Einstein operator (\ref{lin_ein}).
This equation contains higher derivative interactions. 
Just as in the case of scalars one can remove them by considering the
following transformation
\be \label{5dmet}
G_{\m \n} = h^{0}_{\m \n} 
- \frac{1}{12} \left(
\frac{2}{9} D_{\m} D^{\r} \hat{s}^2 D_{\n} D_{\r} \hat{s}^2 
- \frac{10}{3} \hat{s}^2 D_{\m} D_{\n} \hat{s}^2 +
(\frac{10}{9} (D \hat{s}^2)^2 - \frac{32}{9} (\hat{s}^2)^2) g^{o}_{\m
  \n} \right ),
\ee
as can be verified using (\ref{lact}). In terms of these variables 
the field equation becomes
\be \label{5dEin}
R_{\m \n}[G] = 2 (T_{\m \n} - \frac{1}{3} G_{\m \n} T^\s_\s)
\ee
where $R_{\m \n}$ is the Ricci tensor of $G_{\m \n}$ and 
\be
T_{\m \n}= \pa_\m S^2 \pa_\n S^2 - G_{\m \n} (\frac{1}{2} 
(\pa S^2)^2 + V(S^2)).
\ee
The equation (\ref{5dEin}) is indeed the field equation
for $G_{\m \n}$ derived from (\ref{5daction}) and $T_{\m \n}$ 
is the corresponding matter stress energy tensor, where 
keeping with our approximations we only retain terms quadratic
in $S^2$.

Now let us briefly describe the equations determining the KK
gravitons. The traceless part of the ten-dimensional field is given by
\be
h^{I_1}_{(\m \n )} = \phi^{I_1}_{(\m \n )} + \psi^{I_1}_{(\m \n)} +
\frac{1}{(k+1)(k+3)} D_{(\m } D_{\n )} (\frac{2}{5} \pi^{I_1} - 12
b^{I_1}),
\ee
where all fields are the appropriate combinations which are gauge invariant to
quadratic order. $\phi^{I_1}_{(\m \n)}$ is transversal but
$\psi^{I_1}_{\m \n}$ is not; indeed its defining equation is
\be
D^{\m} D^{\n} \psi^{I_1}_{\m \n} = Z^{I_1}[s],
\ee
where $Z^{I_1}[s]$ is quadratic in the field $s^2$ and is determined by the
quadratic corrections to the trace of the Einstein equation in the AdS
directions. The equation for the transversal field is then
\bea
(\Box + 2 - k(k+4) ) \phi^{I_1}_{(\m\n) } &=& (2 L_{E} +8 + k(k+4))
\psi^{I_1}_{ (\m \n )} + Z^{I_1}_{ (\m \n) }; \\
&=& (2 L_{E} + 8 + k(k+4)) \psi^{(t) I_1}_{ (\m \n )}, 
\nn
\eea
where $Z^{I_1}_{ (\m \n)}$ is again quadratic in the field $s^2$ and
follows from the corrections to the $(\m \n)$ Einstein
equation. $\psi^{(t) I_1}_{ (\m \n )}$ is a transversal field which is
quadratic in $s^2$ and contains up to six derivatives. 
We have verified that that the right hand side of the equation can be
written in the latter form; this follows from the detailed structure of
$Z^{I_1}_{ (\m \n)}$ and $\psi^{I_1}_{\m \n}$. It is then immediately
manifest that if one removes the higher derivative terms in the
equation by defining the five dimensional field as 
$\Phi^{I_1}_{\m \n} = \phi^{I_1}_{(\m\n) } - \psi^{(t) I_1}_{ (\m \n
  )}$ then this five dimensional field satisfies the free field equation. 
This is in agreement with the result of \cite{Arutyunov:1999en} which
found the corresponding cubic coupling to vanish and implies
that the five-dimensional field must vanish to the order to which we work.
As we discuss later, there is no physical content in these fields
(to the order to which we work), so we suppress explicit details of the 
(rather complicated) KK reduction map.

\section{Holographic 1-point functions}

\subsection{Generalities}

The KK reduction discussed in the previous sections provides an explicit 
map between ten dimensional solutions and five dimensional
solutions as well as an associated five dimensional action
for gravity coupled to massless and massive KK modes. 
If one would consider this problem in full generality 
the resulting action would involve an infinity of fields.
For determining the holographic 1-point functions, however,
we are only interested in the near boundary behavior of 
the solutions (as is reviewed below). 
The near boundary expansion
effectively decouples all but a finite number of fields,
the number of which depends on the dimension of the operator
whose 1-point function one is computing; the higher the dimension,
the greater the number of fields switched on. 

Starting from a five dimensional action there is a
well developed method for computing holographic
1-point functions, namely holographic renormalization
\cite{dHSS,BFS2}, see \cite{Skenderis:2002wp} 
for a review.
Recall that the basic formula expressing the AdS/CFT 
correspondence \cite{GKP,W} relates the on-shell supergravity 
action with prescribed boundary conditions for all fields
to the generating functional of QFT correlators with the 
boundary fields playing the role of sources\footnote{We work with 
Euclidean signature.}, 
\be \label{GKPW}
\< \exp\left(
-S_{{\rm QFT}}[G_{(0)}] - \int d^4 x \sqrt{G_{(0)}} \cao(x) \F_{(0)}(x)
\right) = \exp(-S_{SG}[G_{(0)},\F_{(0)}])
\ee 
where $G_{(0)}, \F_{(0)}$ are the fields parameterizing the boundary values
of the bulk metric $G$ and of other bulk fields denoted collecting 
by $\F$.

Naively, both sides of this relation diverge: the l.h.s. suffers from 
(the well known QFT) UV divergences and the r.h.s. suffers from IR divergences 
(due to the infinite volume of the spacetime). The divergences on the l.h.s
may be dealt with by standard renormalization. The infinities on the 
r.h.s. are dealt by holographic renormalization. Namely, one
adds a number of boundary counterterms that cancel all possible
infinities that can arise in the on-shell action. Holographic 1-point functions
in the presence of sources are then obtained by 
computing in full generality the first variation of the renormalized 
on-shell supergravity action. This leads to relations
between the 1-point functions and certain coefficients in the near-boundary 
expansion of the bulk fields. This relation effectively 
replaces (\ref{GKPW}) since higher 
point functions can be computing by further differentiating the 1-point 
functions w.r.t. sources. 

The near-boundary expansion of the bulk metric $G_{\m \n}$ and 
scalar field $\Phi^k$, where $k$ is the dimension of the dual 
operator, take the form
\bea \label{near-bdry}
ds_5^2 &=& 
\frac{dz^2}{z^2} 
+ \frac{1}{z^2}\left(G_{(0)ij}(x) + z^2 G_{(2)ij}(x) 
+ z^{4} (G_{(4)ij}(x) + \log z^2 h_{(4)ij}(x))\right) dx^i dx^j; 
\nn \\ 
\F^2(x,z) &=& 
z^2 \left(\log z^2 \F^2_{(0)}(x) + \tilde{\F}_{(0)}^2(x) + \cdots \right); 
\nn \\
\F^k(x,z) &=& z^{(4-k)} \F^k_{(0)}(x) + \cdots +z^k \F_{(2k-4)}^k(x) + \cdots,
\qquad k>2. 
\eea
In these expressions the boundary fields
$G_{(0)ij}, \F^2_{(0)}, \F^k_{(0)}$ parametrize the Dirichlet boundary 
conditions and are also the field theory sources for the 
QFT stress energy tensor and operators of dimension 2 and $k$, 
respectively. The near-boundary
analysis determines all coefficients in these expansions except the 
ones corresponding to the normalizable modes, namely 
$G_{(4)ij}, \tilde{\F}^2_{(0)}, \F^k_{(2k-4)}$. These are related to 
1-point functions, as we review below.

\subsubsection{Radial Hamiltonian formalism}

The structure of the 1-point functions is most transparent 
in the radial Hamiltonian formalism 
\cite{Papadimitriou:2004ap,Papadimitriou:2004rz} (see 
\cite{Kraus:1999di,deBoer:1999xf,Martelli:2002sp} for earlier work). 
So before giving the explicit relation between the 
1-point functions and the coefficients of the 
asymptotic solutions we digress to explain this relation.
Let us define a radial canonical momentum for each field as
\be
\pi=\frac{\partial L}{\partial \Phi'}
\ee
where $L$ is the Lagrangian and prime denotes differential w.r.t.
$r = - \log z$. A covariant 
version of the near boundary expansion in (\ref{near-bdry}) 
is provided by the expansion of momenta in eigenfunctions
of the dilatation operator,
\be
\d_D = \int_{z=\e} d^d x \left( 2 \g_{ij} \frac{\d}{\d \g_{ij}}
+ \sum_k (k-4) \F^k \frac{\d}{\d \F^k} \right) 
= -z \frac{\pa}{\pa z} (1 + O(z)),
\ee
where $\g_{ij}$ is the induced metric at the regulating surface
$z=\e$. The last equality follows
from the leading asymptotics in (\ref{near-bdry}). The near boundary 
expansion (\ref{near-bdry}) now translates into the following expansions
\bea \label{mom_exp}
\pi^i_j(x,\e)&=&\sqrt{\g} (\pi_{(0)}{}^i_j + \cdots + 
\pi_{(4)}{}^i_j + \tilde{\pi}_{(4)}{}^i_j \log \e^2  +\cdots); \nn \\
\pi^k(x,\e) &=& \sqrt{g}
(\pi_{(4-k)}^k + \cdots +\pi_{(k)}^k + \tilde{\pi}^k_{(k)} \log \e^2   
+ \cdots), \qquad k \geq 2
\eea
where $\pi^i_j$ and $\pi^k$ are the radial momenta for the bulk metric  
and the scalar field $\F^k$, respectively, and each term in this expansion 
transforms as indicated by its index
\be
\d_D \pi_{(n)} = - n \pi_{(n)}, 
 \ee
except for $\p_{(4)}{}^i_j, \pi_{(2)}^2, \pi_{(k)}^k$
which transform inhomogeneously, with the inhomogeneous
term being equal to minus two times the coefficient of the logarithmic
term, namely
\be
\d_D \pi_{(4)}{}^i_j = - 4 \pi_{(4)}{}^i_j - 2 \tilde{\pi}_{(4)}{}^i_j,
\qquad \d_D \p^k_{(k)}= - k \p^k_{(k)} - 2  \tilde{\pi}_{(k)}^k.
\ee
One advantage of the momenta expansion (\ref{mom_exp}) over the near boundary 
expansion of the bulk fields (\ref{near-bdry}) is that the 
momentum coefficients $\p_{(n)}{}^i_j, \p_{(n)}^k$ are covariant 
w.r.t. 4d diffeomorphisms that 
respect the regulating hypersurface $z=\e$ whereas the coefficients
in (\ref{near-bdry}) are not.

The coefficients in the momentum expansions can be obtained by 
inserting the expansions in Hamilton's equations. This leads to a
number of equations obtained by collecting all 
terms with the same weight. One then solves these equations 
iteratively and each of them algebraically determines one of the 
coefficients in the expansion in terms of coefficients with lower weight.
This determines all coefficients except $\pi_{(4)}{}^i_j$ and 
$\p^k_{(k)}$, just as the asymptotic 
analysis of the bulk equations
determines all coefficients in (\ref{near-bdry}) except for the 
normalizable modes. 

The renormalized 1-point functions are now simply given  by
the coefficient of the right dimension
\bea
\< T_{ij} \> &=& \pi_{(4)ij} \nn \\
\< \cao^{k} \> &=& \pi^k_{(k)} \label{kop}
\eea
Following \cite{Papadimitriou:2004ap} one can show that there is
a one to one correspondence between the momentum coefficients 
and the coefficients in (\ref{near-bdry}). In particular,
\bea \label{mom1}
\pi^2_{(2)} = \frac{N^2}{2 \pi^2} \left(2 \tilde{\f}_{(0)}\right), \qquad
\pi^k_{(k)} = \frac{N^2}{2 \pi^2} \left((2 k - 4) \f_{(2 k -4)} 
+ {\rm lower}\right)
\eea
where the factor $N^2/2 \pi^2$ is due to the overall factor in 
(\ref{5daction}) and
``lower'' indicates terms with index less than $(2 k -4)$.
These terms are local functions of the sources so they 
are not important in computation of $n$-point functions for $n > 1$
(they lead to contact terms). They are important in the computation of 
vevs  in cases where the solution describes 
a deformation flow \cite{BFS1,Karch:2005ms}.
The specific example we discuss in this paper however is a vev flow 
so we need not specify them.

\subsection{$5d$ supergravity fields}

The part of the $5d$ action involving the metric and the field $S^2$ 
is same as the sector of gauged supergravity analyzed in  
\cite{BFS1,BFS2} (where $S^2$ was called $\F$) 
(see also \cite{Martelli:2002sp,Papadimitriou:2004rz}).
Thus, the results for the 1-point functions carry over unchanged.
The result for $\<\cao^2\>$ is as given above
\be \la{vev2}
\< \cao^2 \> = \frac{N^2}{2 \pi^2} \left(2 \tilde{S}_{(0)}^2\right),
\ee
and for the stress energy tensor 
\bea \label{tij}
\< T_{ij} \> &=& \frac{N^2}{2 \pi^2}\left(
G_{(4)ij} +\frac{1}{3} \tilde{S}_{(0)}^2 G_{(0)ij}+
\frac{1}{8}[\Tr G_{(2)}^2 -(\Tr G_{(2)})^2] G_{(0)ij}\right. \\
&-& \left.\frac{1}{2} (G_{(2)}^2)_{ij} + \frac{1}{4} G_{(2)ij} \Tr G_{(2)}
+\frac{3}{2} h_{(4)ij} 
+(\frac{2}{3} S^2_{(0)} - \tilde{S}_{(0)}^2)
S^2_{(0)} G_{(0)ij}.\right). \nn
\eea

\subsection{KK modes}

We now move to the one point functions of the other fields
$S^4, S^6, T^0, T^2, T^6, \F_{(t)}^2$.
From the last row of Table 1 we see that the 
cubic couplings (relevant for us) vanish for all of them
so to the order to which we are working their field equations 
are just free field equations. One would therefore 
naively conclude that 
the one-point functions are simply given by (\ref{kop})-(\ref{mom1}). 
It turns out however that this is not true and there is 
an additional subtlety. 

The 1-point functions (\ref{kop})
were derived starting from a $5d$ action but in principle one 
should really evaluate on-shell the $10d$ action. In the majority 
of cases this distinction does not make any difference in practice, 
but there is a distinguished
class of additional finite (non-local) boundary terms that one obtains 
from reducing the 10d action. These boundary terms are relevant for the 
computation of extremal correlators, namely $n$-point functions 
of 1/2 BPS operators whose dimensions are 
$\{k_1 {=} k_2 + \cdots + k_n, k_2, \ldots, k_n\}$. 
These correlators are special in that 
they factorize into a product of 2-point functions.

The bulk couplings associated with all extremal 3-point functions 
were shown to be zero in \cite{LMRS} (a result we reproduced 
here for the coupling $(S^2)^2 S^4$, see Table 1). 
The 3-point functions of the corresponding operators however 
are non-zero. It was shown in  \cite{D'Hoker:1999ea} in 
a specific example involving the dilaton and the $t$-field
that even though the bulk contribution to the three point function 
vanishes, there
are additional boundary contributions which 
lead to the correct 3-point function. By supersymmetry,
the same should apply to all other extremal 3-point functions.
It was further conjectured that these results generalize to 
all extremal $n$-point correlators. We refer to \cite{D'Hoker:2002aw} 
for further discussion and references on this issue.

These results should follow from holographic renormalization 
by starting from the $10d$ 
action, requiring that the variational problem is well posed 
and then KK reducing the action with boundary terms. 
Recall that in the $5d$ context all boundary terms, 
including counterterms, are uniquely fixed by the 
requirement that the variational problem is well posed with  
chosen boundary conditions \cite{Papadimitriou:2005ii}. 
We leave a detailed derivation for future work. Here we will
instead use known results in order to fix the form of 
the 1-point functions.

Covariance under $4d$ diffeomorphisms implies that the 1-point function 
should be a function of the coefficients in the momentum 
expansion (\ref{mom_exp}). Furthermore the dimensions should match 
between the l.h.s. and the r.h.s. and the result for 
extremal $n$-point functions should factorize into products
of 2-point functions. This uniquely fixes the form 
of the 1-point functions. For concreteness we discuss
the 1-point functions of $\cao^4$ and $\cao^6$ but the generalization
is obvious. Thus
\bea \label{1pt}
\< \cao^{4} \> &=& 
\pi_{(4)}^4 + a_{422} (\pi^2_{(2)})^2 \\
\< \cao^{6} \> &=& 
\p_{(6)}^6+ a_{642} \pi^4_{(4)} \pi^2_{(2)} + a_{633} (\pi_{(3)}^3)^2 
+a_{6222} (\pi^2_{(2)})^3 \label{o6}
\eea  
where $ a_{422}, a_{642}, a_{633}, a_{6222}$ are numerical constants that 
we show how to compute in the next subsection. Let us explain the 
structure of these 
1-point functions. The leading term $\pi_{(k)}^k$ is the term discussed
above in (\ref{kop}). Observe that the non-linear terms are possible
only when the weight of the operator in l.h.s. can be written as
a sum of weights of other operators, which is also the condition for extremal
correlators. One could also  
consider adding terms involving $\pi_{{n}}^k$ with $n<k$. 
Such terms can only possibly contribute to the coincident limit
of correlators (since $\pi_{{n}}^k$ with $n<k$ is local in the sources) 
or to vevs of solutions describing deformation flows, 
so they are not important for our analysis.
 
\subsection{Extremal couplings}

In this subsection we will compute the coefficients $a_{422}, a_{6222}$.
The coefficients $a_{633}$ and $a_{642}$ could be computed in similar way
but we will not need their explicit values in this paper. To obtain
the coefficients  $a_{422}, a_{6222}$ we compute the 
3- and 4-point functions starting from (\ref{1pt}) and then fix
the coefficients so that the numerical factors agree with the 
the computation in free field theory. Note that the dependence
on the coordinates is guaranteed to be correct by the structure 
of (\ref{1pt}) (as should become clear shortly).

We start with the computation of $a_{422}$.
By definition
\bea 
\< \cao^{k}(x) \cao^{k}(y) \> &=& 
-\left.\frac{\delta \< \cao^{k}(x)\> }{\d \f^k_{(0)}(y)}\right|_{\f^k_{(0)}=0}=
\left. 
- \frac{\delta \pi^k_{(k)}(x)}{\d \f^k_{(0)}(y)}
\right|_{\f^k_{(0)}=0} \label{kk} \\
\< \cao^{4}(x_1) \cao^{2}(x_2) \cao^2(x_3) \> &=& 
\left. 
\frac{\delta^2 \< \cao^4(x_1)\>}{\d \f^2_{(0)}(x_2) \d \f^2_{(0)}(x_3)} 
\right|_{\f^2_{(0)}=0} \label{422} \\
&&\hspace{-2cm}=
\left. \left(
\frac{\delta^2 \pi^4_{(4)}(x_1)}{\d \f^2_{(0)}(x_2) \d \f^2_{(0)}(x_3)}
+ 2 a_{422} \left(\frac{\delta \pi^2_{(2)}(x_1)}{\d \f^2_{(0)}(x_2)}\right)
\left(\frac{\delta \pi^2_{(2)}(x_1)}{\d \f^2_{(0)}(x_3)}\right)\right)
\right|_{\f^2_{(0)}=0} \nn
\eea
To evaluate these expressions we need to know $\pi^k_{(k)}$ 
to linear order in  $\f_{(0)}^k$ 
and $\pi^4_{(4)}$ to quadratic order in $\f_{(0)}^2$.
Recall that $\pi_{(k)}^k$ is proportional to the vev part of the solution
(see (\ref{mom1})), so in order to compute 2- and 3-point functions 
we need to solve bulk equations expanded around the 
background solution (which in the current context is just $AdS$)
to linear and quadratic order, respectively, and then extract the 
coefficient of order $z^{k}$. 
For extremal couplings the cubic coupling 
to $S^2$  is zero, $\l_{422}=0$, so the bulk equation for $S^4$ 
continues to be $\Box S^4 =0$ and the solution does not acquire dependence
on $\f_{(0)}^2$ so the second variation of 
$\pi^{(4)}_4$ w.r.t. $\f_{(0)}^2$ is zero\footnote{In the non-extremal
case, the bulk equation reads $\Box S^k=\l_{klm} S^l S^m$ and the
r.h.s. induces a correction to  $S^k$ proportional to
$z^k (\f_{(0)}^l \f_{(0)}^m)$ so the second variation of 
$\pi^{(k)}_k$ w.r.t. $\f_{(0)}^l$ and $\f_{(0)}^m$ is non-zero
yielding the 3-point function, see \cite{Skenderis:2002wp} 
for a detailed discussion.}.  Therefore only the last term 
in (\ref{422}) contributes and using (\ref{kk}) we see that
the extremal
3-point function is a product of 2-point functions
\be \label{su_3pt}
\< \cao^{4}(x_1) \cao^{2}(x_2) \cao^2(x_3) \>=2 a_{422} 
\< \cao^{2}(x_1) \cao^{2}(x_2) \>
\< \cao^{2}(x_1) \cao^{2}(x_3) \>
\ee
Since this 3-point function does not renormalize one can compute it 
via free fields, which allows us to fix the proportionality constant 
$a_{422}$.

In the large $N$ limit (i.e. dropping non-planar contributions) 
the free field computations for the 2- and (extremal) 3-point functions yield
\bea
\left < \cao^{k}(x) \cao^{k}(y) \right > &=&
{\cal N}_{k}^2 \l^k \frac{k}{(2 \pi)^{2k} 
\left | x - y \right  |^{2k}}, \label{2pt}\\
\left < \cao^{k_1}(x_1) \cao^{k_2}(x_2) \cao^{k_3}(x_3) \right > &=&
{\cal N}_{k_1} {\cal N}_{k_2} {\cal N}_{k_3}  \frac{\l^{k_1}}{N}
\frac{k_1 k_2 k_3\< C_{k_1} C_{k_2} C_{k_3} \>}{ (2 \pi)^{2 k_1} \left | x_1 - x_2 \right |^{k_1} 
\left  | x_1 - x_3 \right |^{k_1} }, \label{3pt}
\eea
where the operators are defined in (\ref{CPO}) and in the extremal 3-point
function $k_1 = k_2 + k_3$. The normalization
factors ${\cal N}_{k}$ are chosen such that the 2-point function
of (\ref{2pt}) agrees with the supergravity results; in particular
given that for $k \neq 2$ the supergravity result is 
\be
\left < \cao^{I_1}(x) \cao^{I_2}(y) \right > = \frac{N^2}{2 \pi^2}
\left (\frac{ \G(k+1)}{\pi^2 \G(k-2)} \frac{(2k-4)}{k} \frac{\d^{I_1 I_2}}{
  \left | x - y \right |^{2k}} \right ),
\ee
whilst for $k=2$ the result is 
\be
\left < \cao^{2}(x) \cao^{2}(y) \right > = \frac{N^2}{2 \pi^2}
\left (\frac{2\d^{I_1 I_2}}{\pi^2
  \left | x - y \right |^{4}} \right).
\ee
The normalizations are thus
\bea \label{norm}
{\cal N}_{I_1} &=& \frac{N}{\l^{\half k}} \frac{ 2^k \pi^{k-2}}{k}
\sqrt{ \frac{\Gamma(k+1) (2k-4)}{2 \Gamma(k-2)}}; \qquad k \neq 2, \\
{\cal N}_{2} &=& 2 \sqrt{2} \frac{N} {\l}. \nn 
\eea
 Inserting 
(\ref{2pt}) in (\ref{su_3pt}) and comparing with (\ref{3pt})
leads to
\be
a_{4 2 2} = \frac{2 {\cal N}_{4}}{{\cal N}_{2}^2 N} \< C_4 C_2 C_2 \> 
= \frac{3 {\cal N}_{4}}{\sqrt{5} {\cal N}_{2}^2 N}, 
\ee
where in the last equality we have used the explicit value of 
the triple overlap $\< C_4 C_2 C_2 \>$. This can be obtained from the formulae 
(\ref{3over})-(\ref{y_norm})
by computing the overlap of $Y^{(4,0)}$ and $(Y^{(2,0)})^2$
using the explicit expressions (\ref{y_expl}).

The computation of the coupling $a_{6222}$ is analogous.
The bulk quartic coupling was shown to be zero in \cite{Arutyunov:1999fb}
and thus the only contribution to the
extremal 4-point function comes from the last term in (\ref{o6})
\be
\<\cao^6(x_1) \cao^2(x_2)  \cao^2(x_3)  \cao^2(x_4) \>
= 6 a_{64222} \prod_{k=2}^4 \< \cao^{2}(x_1) \cao^{2}(x_k) \>
\ee
The free field result for extremal 4-point functions (in the
planar limit) is
\be
\< \cao^{k_1}(x_1) \cao^{k_2}(x_2) \cao^{k_3}(x_3) \cao^{k_4}(x_4) \> =
(\prod_{i} {\cal N}_{i}) \frac{\l^{k_1}}{N}
\frac{k_1 k_2 k_3 k_4 \< C_{k_1} C_{k_2} C_{k_3} C_{k_4} \> }
{ (2 \pi)^{2 k_1} \left | x_1 - x_2 \right |^{2k_2} 
\left  | x_1 - x_3 \right |^{2k_3} \left  | x_1 - x_4 \right |^{2k_4}}, \label{4pt}
\ee
where $k_1 = k_2 + k_3 +k_4$, which fixes the proportionality constant to be
\be
a_{6222} = \frac{{\cal N}_{6}}{{\cal N}_{2}^3 N} \< C^4 C^2 C^2 C^2\> 
= \frac{{\cal N}_{6}}{{\cal N}_{2}^3 N} \frac{3 \sqrt{3}}{5 \sqrt{7}}
\ee
where $\< C^4 C^2 C^2 C^2\>$ was computed using the following integral
formula valid for extremal overlaps
\be
\int Y^{I_1} Y^{I_2} Y^{I_3} Y^{I_4}=
\frac{\pi^3}{(k_1+1)(k_1+2) 2^{k_1-1}}  
\< C^{I_1} C^{I_2} C^{I_3} C^{I_3}\>,
\ee
along with the explicit forms of the spherical harmonics.

\subsubsection{Summary}

To summarize we have shown that 1-point functions 
of the operators $\cao^4, \cao^6$ are \footnote{The non-linear terms
in these relations may have the interpretation as operator
mixing between single trace and multi-trace operators.} 
\bea 
\< \cao^{4} \> &=& 
\pi_{(4)}^4 + \frac{3 {\cal N}_{4}}{\sqrt{5} {\cal N}_{2}^2 N}
(\pi^2_{(2)})^2 \label{1ptfun} \\
\< \cao^{6} \> &=& 
\p_{(6)}^6+ a_{642} \pi^4_{(4)} \pi^2_{(2)} + a_{633} (\pi_{(3)}^3)^2 
+ \frac{{\cal N}_{6}}{{\cal N}_{2}^3 N} \frac{3 \sqrt{3}}{5 \sqrt{7}}
(\pi^2_{(2)})^3 \nn
\eea  
Notice that although we have used the 3- and 4-point functions on $AdS$ to 
fix the couplings $a_{422}$ and $a_{6222}$, these 1-point functions
hold for {\it any} solution of the bulk field equations.

Notice also that these 1-point functions are compatible with the 
(conjectured) structure of near-extremal correlators.
Recall that 
near-extremal correlators have weights $k_1=k_2 + \cdots k_n -2 m$
with $0 \leq m \leq n-2$. These correlators are conjectured 
(and checked through order $g^2$) to
be sums of terms each of which factors into products of 
lower-point correlators \cite{D'Hoker:2000dm}. When $m=1$ the 
correlator is called next-to-extremal and factorizes into a 3-point function 
and $(n{-}2)$ 2-point functions. One can easily check that this 
structure emerges from (\ref{1ptfun}) after using the fact that
the bulk coupling vanishes. (To be more precise, the bulk coupling 
is known to vanish up to quartic order and is conjectured to vanish to
all orders). For example 
the next-to-extremal correlator $\<\cao^4 \cao^2 \cao^2 \cao^2\>$ 
is given by
\bea
&&\< \cao^{4}(x_1) \cao^{2}(x_2) \cao^{2}(x_3) \cao^{2}(x_4) \> =
2 a_{422}
\left(\frac{\d \pi^2_{(2)}(x_1)}{\delta \f_{(0)}^2(x_2) 
\delta \f_{(0)}^2(x_3)}\right)  
\left(\frac{\d \pi^2_{(2)}(x_1)}{\delta \f_{(0)}^2(x_4)}\right) + \cdots \nn \\
&&\qquad =2 a_{422}
\left(\<\cao^4(x_1) \cao^2(x_2) \cao^2(x_3)\> \<\cao^2(x_1) \cao^2(x_4)\>
+\cdots\right)
\eea
where the dots indicate permutation in $x_2, x_3, x_4$.

\section{Coulomb branch solution}

\subsection{Continuous distributions of D3 branes}

It is intuitively clear that the Coulomb branch of ${\cal N}=4$ SYM should be 
described by multi-center D3 brane solutions. Solutions describing N separated 
D3 branes solve the field equations that follow from 
the bulk supergravity action  
coupled to the worldvolume action of N (separated) D3-branes. 
It will be important for us 
to keep track of all normalizations factors so we set the stage by first
reviewing some standard material.
The bulk action is normalized as
\be \la{bulk10d}
S = \frac{1}{2 \k^2} \int d^{10}x \sqrt{-g} (R + \cdots ), 
\qquad 2 \k^2 = (2 \pi)^7 (\a')^4 g_s^2,
\ee
and the worldvolume theory is given by
\be \la{source}
S_{{\rm source}} = \sum_{a=1}^{N} \int d^{10}x \int d^{4}\s^a
{\cal L}_{DBI}(\s^a) \d(x^M - X^M(\s^a)),
\ee
where the Lagrangian for each D-brane is normalized as
\be
{\cal L}_{DBI}(\s^a) = T_{3} (\sqrt{{\rm{det}} (\g + 2 \pi \a' F)} + \cdots),
\qquad T_3 = \frac{1}{(2 \pi)^3 (\a')^2 g_s},
\ee
where $\g_{ij} = \pa_{i} X^{M} \pa_{j} X^{N} g_{MN}$ 
is the induced metric and derivatives
are with respect to the worldvolume coordinates $\s^i$.
The D3-brane solutions take the form
\bea
ds^2 &=& H(x_\perp) ^{-1/2} d x_{||}^2
+ H(x_\perp)^{1/2} d x_\perp^2 \\
F&=&\frac{1}{4} (d H^{-1} \wedge \omega_{||}
- *_\perp d_\perp H)
\eea
where $\omega_{||}$ is the volume form in the worldvolume directions, 
$*_\perp$ and $d_\perp$ refer to the Hodge star and exterior derivative
in the flat overall transverse directions and $H$ is a harmonic
function. We are interested in the case of a uniform distribution of N
D3-branes on a two dimensional disc of radius $l$. Approximating
the distribution as a continuum distribution with density
\be
\rho(r) = \frac{N}{\pi l^2} \theta(l^2 - r^2),
\ee
where $r$ is the radial coordinate in the plane of the distribution,
the solution for the harmonic function is (see for example 
\cite{Freedman:1999gk,Brandhuber:1999jr})
\be
H=\frac{L^4}{\pi l^2} \int_{r'\leq l} d^2 r' 
\frac{1}{(\vec{x}_\perp - \vec{r'})^4} 
= -\frac{L^4}{2 l^2 y^2} 
\left(\frac{r^2 - l^2 + y^2}{\sqrt{(r^2 + l^2 +y^2)^2-4 r^2 l^2}} -1 \right)
\ee
where $L^4 = 4 \pi g_s N (\a')^2$, 
$\vec{y}$ are coordinates in the four dimensional space
transverse to the distribution of the D3 branes and $\vec{r}$ lies
in the plane of the distribution.

A change of coordinates
\be
y = \bar{r} \sin \bar{\q}, \qquad r = \sqrt{l^2 + \bar{r}^2} \cos \bar{\q}
\ee
brings the solution into the form
\bea
ds^2 &=& 
\frac{\bar{r}^2 \z}{L^2} \left(dx_{||}^2 + \frac{L^4 d \bar{r}^2}{\bar{r}^4 \l^6}\right)
+ \frac{L^2}{\z} \left(\z^2 d \bar{\q}^2 
+ \sin^2 \bar{\q} d \W^2 + \l^6 \cos^2 \bar{\q} d \f^2\right) \\
F&=& L^{-4} \left(\bar{r}^3(1 + \frac{l^2}{2 \bar{r}^2} \sin^2 \bar{\q}) 
d \bar{r}
+ \frac{1}{4} l^2 \bar{r}^2 \sin 2 \bar{\q} d \bar{\q} \right)
\wedge \omega_{||}  \nn \\
&+& L^4 \sin^3 \bar{\q} \cos \bar{\q} \frac{1}{\z^4}
\left(\l^6 (1 + \frac{l^2}{2 \bar{r}^2} \sin^2 \bar{\q})
d \bar{\q} - 
\frac{l^2}{4 \bar{r}^3} \sin 2 \bar{\q} d \bar{r}\right) 
\wedge d \W_3 \wedge d \f \nn
\eea
where 
\be
\z^2 =1 + \frac{l^2}{\bar{r}^2} \sin^2 \bar{\q}, \qquad \l^6 = 1 
+ \frac{l^2}{\bar{r}^2}
\ee
Now note that if we rescale the four dimensional coordinates as
\be \la{scaling}
x_{||} \rightarrow L^2 x_{||}
\ee
the metric has an overall $L^2$ factor
whilst the five form has an overall $L^4$ factor. These factors
combined with the prefactor of (\ref{bulk10d}) result in the overall
normalization factor of the five dimensional action (\ref{5daction});
we can therefore suppress the $L$ prefactors in the rest of this
section. 

\subsection{Asymptotic expansion}

We now wish to expand the metric near the boundary. A
systematic way to do this is to use Gaussian normal
coordinates centered at the boundary of $AdS_5$ 
and then expand all fields using the radial coordinate
as a small parameter. This radial axial gauge can be reached by the 
charge of coordinates
\bea
\frac{l}{\bar{r}} &=& z (1 + a_1 z^2 + a_2 z^4 + O[z]^6) \nn \\
\bar{\q} &=& \q + b_1 z^2  + b_2 z^4 + O[z]^6 
\eea
where
\bea
a_1 &=& \frac{1}{2^3} (2 - \sin^2 \q), \quad 
a_2 =\frac{1}{2^8} \cos^2 \q(5 + 11 \cos 2\q) \nn \\
b_1 &=& \frac{1}{2^3} \cos \q \sin \q, \quad 
b_2 =\frac{5}{2^{9}} (-\sin 2\q + \sin 4\q)
\eea
The metric then takes the form
\bea
ds^2 &=& \frac{dz^2}{z^2} + 
\frac{l^2}{z^2}(1+ \a_1 z^2 + \a_2 z^4) d x_{||}^2 \\
&&+ d\q^2 (1+\b_1 z^2 + \b_2 z^4) + \sin^2 \q d \W_3^2 
(1 + \g_1 z^2 + \g_2 z^4) + \cos^2 \q d \f^2 (1 + \d_1 z^2 
+ \d_2 z^4) 
\nn
\eea
where the coefficients $\a_1, \a_2$ etc. depend on the angular
coordinate $\theta$. By scaling 
\be
z  \to z l
\ee
the leading metric becomes a unit radius $AdS_5 \times S^5$
and factors of $l$ appear in the fluctuations. These factors can be
easily reinstated in the final formulae so for simplicity we set $l=1$
for now. 
Using the explicit form of spherical harmonics in (\ref{y_expl})
and (\ref{t_harm}) the deviation of the metric from $AdS_5 \times S^5$ 
can be rewritten as in (\ref{fluct_h}) with the following
coefficients (valid up to terms of order $z^4$),
\bea
&& \td{h}_{ij}^0(z) = \frac{1}{32} z^2 \d_{ij}, \quad \td{h}_{ij}^2(z) = 
\sqrt{12} \left(-\frac{1}{4} + \frac{23}{320} z^2 \right) \d_{ij}, \quad
\td{h}_{ij}^4(z) = -\frac{3 \sqrt{5}}{20} z^2 \d_{ij}, \nn \\
&&\hat{\phi}^2_{(t)}(z) =  \frac{3}{160} z^4, \quad 
{\phi}^2_{(s)}(z)= \sqrt{12} (\frac{1}{8} z^2 -\frac{1}{256} z^4), \quad 
{\phi}^4_{(s)}(z)= \frac{\sqrt{5}}{32} z^4 \label{metric_cor}\\
&& \td{\pi}^0(z)=\frac{1}{8} z^4, \quad \td{\pi}^2(z)= \sqrt{12} 
(z^2 -\frac{17}{64} z^4), \quad  
\td{\pi}^4(z)= \frac{3 \sqrt{5}}{2} z^4. \nn
\eea
Similarly, from the expansion of the five form we obtain
\bea
b^2_{(s)} &=& -\frac{\sqrt{3}}{8} z^2 + \frac{31 \sqrt{3}}{1280} z^4 \\
b^4_{(s)} &=& -\frac{39 \sqrt{5}}{640} z^4  \nn
\eea
There are several comments in order here. Firstly, the solution 
is not in the de Donder gauge, as one can see from the fact that 
the scalar fields $\f_{(s)}^2$ and $\f_{(s)}^4$ are non-zero.
Secondly, the expansion contains many more non-zero fields
that one would naively expect. In particular, there are non-zero
KK gravitons, $\td{h}_{ij}^2$ and $\td{h}_{ij}^4$
(which are dual to the operators of the schematic
form, $\Tr F_+ F_- X^k$ for $k=2,4$ (see Table 7 of \cite{D'Hoker:2002aw})), 
scalar field $\f_{(t)}^2$
(that couples to a tensor harmonic and is dual to the operator 
$\Tr \l \l \bar{\l} \bar{\l} X^2$) scalar fields 
$t^0, t^2, t^4$ (that are dual to $\Tr F_+^2 F_-^2 X^k$ for $k=0,2,4$) 
and scalar fields $s^2, s^4$ (that are dual to the operators $\Tr X^k$, 
$k=2,4$). However, we know that in the CB flow only the operators
$\Tr X^k$ get a vev. So what is the meaning of the values of the 
additional fields?

To answer this question we should apply our map to obtain the 
gauge invariant five dimensional fields. As a first step 
we need to construct gauge invariant combinations. Using 
(\ref{pi1})-(\ref{b1})-(\ref{pi0}) and (\ref{phit}) 
and the definition of $t^k$, $s^k$ we get
\bea
&& t^0 = - \frac{1}{128} z^4, \qquad 
t^2 = - \frac{\sqrt{3}}{160} z^4, \qquad t^4 = - \frac{3 \sqrt{5}}{160} z^4, \\
&& s^2 = \frac{\sqrt{3}}{4} z^2 - \frac{\sqrt{3}}{160} z^4, \qquad 
s^4 = \frac{37 \sqrt{5}}{160} z^4, \qquad \f^2_{(t)} = 0.
\eea
The five-dimensional fields are obtained from these by the KK reduction
formula (\ref{10to5}) yielding
\be \label{5dvalues}
T^0 = T^2 = T^4 = \Phi_{(t)}^2 = 0, \qquad 
S^2 = \frac{1}{\sqrt{6}} (z^2 - \frac{1}{6} z^4), \qquad S^4=0.
\ee
We thus see that that all additional scalar fields are equal to zero! 
The same is also true for the 
KK gravitons but we do not give the details here. We would like 
to emphasize, however, that a non-zero answer
for these fields would not be a problem for the duality. The only 
cases where it would problematic is if the non-zero values
correspond to a source or a vev. Note that 
all additional fields correspond to irrelevant operators
and a non-zero source would not be consistent with 
$AdS$ asymptotics. Furthermore, the corresponding vev part 
would appear at much higher power of $z$. To understand the 
(possibly non-zero) values note that the $5d$ equations are 
schematically of the form
\be
(\Box + m_{\Phi}^2) \Phi = \l_{\Phi22} (S^2)^2 + \cdots 
\ee
where $\Phi$ denotes collectively the fields other than 
$s^2, s^4$ and the metric that are turned on.
Any non-zero value for these fields would simply be induced by 
interaction terms -- such non-zero fields just reflect the 
non-linear structure of gravity. 
In our case, it turns out that the couplings $\l_{\Phi22}$ are zero,
so the fields $\Phi$ had to zero to this order.

The fields that are important to understand at each order 
$z^k$ are the ones which correspond to operators
whose vevs can receive a contribution from the asymptotics at this order. 
In our case, these are the fields $S^2, S^4$ and the metric $G_{\m \n}$.
The solution we discuss can be reduced to five dimensions
using a ``consistent truncation ansatz'' (see 
\cite{Freedman:1999gk,Cvetic:2000nc}). The reduced model 
involves the metric and $S^2$. The expression for $S^2$ 
in (\ref{5dvalues}) exactly agrees with the asymptotic 
expansion of the $5d$ solution, compare with (5.2) of 
\cite{BFS2} (and use $\F = -S^2$). We will return to the 
metric momentarily.  

The expression for $S^4$ in (\ref{5dvalues}) is new information. 
The fact that it is zero comes out of  non-trivial cancellations
and at first sight is surprising since the vev of the dual operator is 
non-zero. From Table 1 we see that the coupling
of $S^4$ to $(S^2)^2$ is zero; this is an example of an 
extremal coupling
\be \label{s4eqn}
\Box S^4 =0.
\ee
In this case however the vanishing of the coupling 
only explains the absence of logarithmic terms in the asymptotic
expansion of $S^4$. Logarithmic terms in the asymptotic expansion
are related to conformal anomalies. Such conformal anomalies
due to 3-point functions are 
possible when the couplings are extremal, see section 
2 of \cite{Petkou:1999fv}. They are however proportional to 
the sources  so they evaluate to zero on the Coulomb branch,
in agreement with the absence of logarithmic terms in the
asymptotic expansion. In other 
words, the vanishing of the extremal couplings is {\it required} by 
the AdS/CFT correspondence and the structure of the 
conformal anomaly.  Equation (\ref{s4eqn}) allows for a homogeneous
solution that is proportional to $z^4$ and one might have 
anticipated that the homogeneous term would be non-zero,
since the vev of the dual operator is 
non-zero. We will resolve this issue in the next subsection. 
  
We now return to the spacetime metric.
We see from (\ref{metric_cor}) that the metric 
is corrected at the normalizable 
mode order. More precisely the combination that diagonalizes the 
field equations to linear order is
$h'_{\mu \nu}= (\td{h}^{0}_{\mu \nu} + \frac{1}{3} \td{\pi}^0 g^{o}_{\mu
  \nu})$ and in our explicit solution this is given by
\be \label{naive5d}
h'_{zz} = \frac{1}{24} z^2, \qquad
h'_{ij} = \frac{7}{96} z^2 \d_{ij}.
\ee
Naively this would imply that the dual stress energy 
tensor is non-zero. The solution, however, is supersymmetric 
so the vev of the stress energy tensor must be equal to zero.
(We discussed this point earlier in the introduction.)
As mentioned above the 10d solution can be reduced to five dimensions
using a consistent truncation ansatz. The asymptotics of 
the $5d$ metric were given in the introduction in (\ref{5d_CB}). 
As mentioned there, despite the non-zero coefficient of the 
$\hat{z}^4$ term, the vev of $T_{ij}$ is zero because of 
additional contributions to the 1-point function. 
This does not immediately resolve the issue however because
the metric in (\ref{naive5d}) does not agree 
with the metric in (\ref{5d_CB}) (when both written in the 
same gauge)! 

The issue here is that  $h'_{\mu \nu}$ is not the correct 
$5d$ metric. Firstly, $h'_{\mu \nu}$
does not transform correctly, i.e. as a five-dimensional metric. 
As derived in section \ref{g2ndt} the
combination which transforms properly is $h^0_{\mu \nu}$
in (\ref{ginv_h0}). Evaluating this formula for the case 
at hand gives
\be
h^{0}_{zz} = - \frac{11}{48} z^2, \qquad
h^{0}_{ij} = \frac{19}{192} z^2 \d_{ij}.
\ee
The five-dimensional metric is now obtained by using the non-linear
KK map in (\ref{5dmet}). The resulting five-dimensional metric
(including the background term) is given by
\be
ds^2 = (1 - \frac{13}{144} z^4) \frac{dz^2}{z^2} 
+ \frac{1}{z^2} (1 - \frac{19}{576} z^4) dx_{||}^2.
\ee
This metric is not in the same gauge as the metric in (\ref{5d_CB}).
To correct for that we change coordinates as
$z = \hat{z} (1 + \frac{13}{1152} \hat{z}^4)$ and the 
metric becomes
\be
ds^2 = \frac{d\hat{z}^2}{\hat{z}^2} + \frac{1}{\hat{z}^2} (1 -
\frac{1}{18} \hat{z}^4)  dx_{||}^2,
\ee
which precisely agrees with (\ref{5d_CB}) 
(after reinstating the factors of l)! Note that this coordinate change
does not affect any other fields to the order we work to.

\subsection{Comparison with field theory}

Given the asymptotic expansions of the five dimensional fields we can
now read off the vevs for the corresponding corresponding operators.
These must agree with the field theory results
discussed in section 2 because, as we explained there, 
${\cal N}=4$ supersymmetry forbids any quantum corrections, 
so this computation is a test of the gravity/gauge theory
duality. 

Using the formula (\ref{vev2}) we see that 
the vev of the $\D = 2$ operator is given in terms of the normalizable
mode $\td{S}^2_0(x)$ of the supergravity field $S^2$ as 
\be \la{sg1}
\left < \cao^{2} \right >_{SUGRA} = \frac{N^2}{2 \pi^2} (2 \td{S}^2_0(x)) =
\frac{N^2}{\sqrt{6} \pi^2} l^2 ,
\ee
where we use the explicit result for our solution $\td{S}^2_0 = 1/\sqrt{6}$
and we reinstated the factors of $l$.
This 1-point function was previously derived in \cite{BFS1}.
To compare with the field theory expression (set $n=1$ in  (\ref{vev1})),
\be \label{qftvev}
\left < \cao^{2} \right >_{QFT} = \frac{{\cal N}_2 a^2}{2 \sqrt{3}} N
\ee
we need to  normalize operators and the action in the same way. 
The normalization of the operators was given in (\ref{norm}).
Expanding the source action (\ref{source}) to leading
order in $\a'$ and taking into account the rescaling of the $R^{3,1}$
coordinates in (\ref{scaling}) the kinetic term for the scalar fields
is normalized as 
\be
T_{3} L^4 \int d^4 \s (\frac{1}{2} (\pa X)^2) = \left (\frac{\l}{2
  \pi^2} \right ) \frac{1}{g_s} \int d^4
\s (\frac{1}{2} (\pa X)^2).
\ee
On the other hand, the field theory computation 
was done with canonically normalized scalars. It follows that the 
radius $a$ of the distribution in field theory (with canonically normalized 
scalars) is related to the radius $l$ of distribution of D3 branes by
\be \la{rad1}
a = \sqrt{\frac{\l}{2 \pi^2}} l.
\ee
Using the normalization ${\cal N}_2$ in (\ref{norm})
and (\ref{rad1}) we find precise agreement between the 
supergravity and field theory computations!

Next consider the $\D=4$ operator.
Given the result of the previous subsection that 
the normalizable mode of the corresponding supergravity 
field vanishes, the vev (\ref{1ptfun}) receives contributions only
from the term quadratic in $\pi^2_{(2)}$:
\be \la{sg2}
\left < \cao^{4} \right > = 
\frac{3 {\cal N}_{4}}{\sqrt{5} {\cal N}_{2}^2 N} (\< \cao^{2} \>)^2
= \frac{{\cal N}_{4} a^4}{2^2 \sqrt{5}} N
\ee
where we used (\ref{kop}) and (\ref{qftvev}).
This is precisely the correct field theory vev!
\bigskip

Now let us consider the $\D = 6$ operator. The vanishing of
$\pi^4_{(4)}$ and $\pi^3_{(3)}$ means that in this case the formula
(\ref{1ptfun}) reduces to just
\be
\left < \cao^{6} \right > = \pi^6_{(6)} + 
\frac{{\cal N}_{6}}{{\cal N}_{2}^3 N} \frac{3 \sqrt{3}}{5 \sqrt{7}}
(\< \cao^{2} \>)^3 
\ee
The latter of these terms evaluates to 
\be \la{poi}
\frac{1}{5} \left(\frac{{\cal N}_{6} a^6}{2^3 \sqrt{7}} N\right) = 
\frac{1}{5} \left < \cao^{6} \right >_{QFT}
\ee
Given that in the $\D = 4$ case, there was no contribution to the vev
from the bulk supergravity field one might have wondered whether the same was
true in the $\D =6$ case, and indicative of a more general result. However,
(\ref{poi}) only accounts for one fifth of the field theory result,
and there must therefore be an additional contribution from the
supergravity field dual to $\cao^6$. To verify this  
one would have to extend our supergravity computations to one order 
higher, including quartic couplings, 
to extract the normalizable mode of the supergravity field $S^6$.
Note that the structure of the 1-point functions
is such that the terms non-linear in momenta always give a contribution
that is proportional to the QFT vev. It is a
curious fact that up to at least $\cao^{10}$ (which is as far as we
explicitly checked)
the proportionality coefficient is a rational number, 
despite the fact that intermediate formulae contain square roots.

The vev of the stress energy tensor was already computed
in \cite{BFS1} (using (\ref{tij})) and, as noted earlier, 
was found to be zero, in agreement with the fact that the solution 
is supersymmetric. We have thus succeeded in showing that the vevs of 
all operators 
up to dimension $\D=4$ are correctly reproduced by supergravity!

\section{Conclusions}

We have developed in this paper a systematic method for constructing 
the holographic map between the asymptotics of a ten dimensional
solution and the 1-point functions of the dual QFT. Our main goal 
was to develop an unambiguous method that can, at least in principle,
always be carried out. 
The main elements entering our construction are (i) the
development of a gauge invariant version of KK reduction;
(ii) construction of the KK map to non-linear order and
(iii) application of holographic renormalization, including a
proper treatment of extremal couplings. 

One-point functions can be derived rigorously starting from 
a 5d action via holographic renormalization. Our strategy 
for obtaining the 1-point functions dual to general KK fields
was thus to reduce the field equations over the compact manifold and then 
use holographic renormalization. Recall that holographic 
renormalization relates the vevs to coefficients 
in the asymptotic expansion of the 5d solution.  So to compute the
vevs starting from a 10d solution one has to understand 
quantitatively how the solution is reduced to five dimensions
at the non-linear level. The point is that non-linear terms
can give a contribution at exactly the same order (in a radial 
expansion) as linear terms. However, 1-point functions
of operators of a given dimension can only receive contribution
from non-linear terms involving fields dual to operators
of lower dimension.

The KK reduction map is constructed by first 
computing the fluctuation equations around
$AdS_5 \times S^5$ to a certain order in the fluctuation 
fields and then finding the field transformation that 
removes the higher derivative terms from these
equations. This field transformation {\it is} the 
KK reduction map (to this order) and the resulting field equations 
{\it are} the 5d field equations. 

We would like to contrast our procedure with the 
procedure of ``consistent truncuation''. In the latter 
one only keeps certain (typically low lying)
modes in the KK reduction and then has to prove that 
the dynamics of these modes decouple from the rest.
The existence of such a truncation is highly non-trivial and only holds for
special compactifications. In the AdS/CFT correspondence
consistent truncation maps to the closure of a subset of operators
of the CFT under OPEs. In our discussion we keep all KK modes,
so there is never an issue of consistency.
We are however interested only in the asymptotic expansion
of the resulting field equations. This effectively decouples all
but a finite number of fields at each order in the expansion. In particular,
for computing the vev of an operator of a given 
dimension only the fields dual to operators of the same or lower dimensions need 
to be kept. 

The need for a gauge invariant KK reduction stems from the fact that
the KK reduction is most efficiently done in a specific gauge, 
the de Donder gauge, but in general 
explicit solutions will not be - and many known
interesting solutions are not - in this gauge. Reaching this gauge
would require finding a transformation that in general 
is not easy to obtain (at the non-linear level). 
Thus instead of gauge fixing the 
diffeomorphisms we construct gauge invariant
variables. This allows us to immediately lift results 
derived in one gauge to another gauge: one can first 
obtain the KK map in the de Donder gauge and then 
relax the gauge condition by simply replacing all fields by their
gauge invariant generalization.
The construction of the gauge invariant variables can be done systematically
in the number of fields and we have done so up to second 
order in the fields.

A final subtlety involves extremal couplings. Extremal correlators
involve operators with the dimension of one of them 
equal to the sum of the dimensions of all the others. Such correlators
are non-zero and are believed not to renormalize.
A naive computation in supergravity however would give zero because the 
corresponding bulk coupling vanishes. It was argued
in \cite{D'Hoker:1999ea} that precisely in these cases
there are additional boundary terms, originating from the 
higher derivative terms in the fluctuation equations, 
that one should take into 
account when evaluating the on-shell action and these yield
the correct answer. In holographic renormalization one 
effectively replaces the on-shell action by renormalized 
1-point functions in the presence of sources. These 
1-point functions are valid for {\it any} solution of the 
field equations and higher $n$-point functions can be computed
by further differentiating w.r.t. sources. Additional
boundary terms in the $5d$ action, beyond the ones implied by 
the bulk 5d action via the variational problem,
would manifest as additional contributions to the renormalized
1-point functions. The form of the additional terms is uniquely
fixed by general principles. This leaves a few numerical coefficients
to be determined and these can be easily computed
by comparing the extremal correlators computed in weakly coupled
${\cal N}=4$ SYM and in supergravity.

Combining these elements one obtains a well-defined holographic 
map. In our discussion
we focused on solutions of IIB supergravity that involve only the metric 
and the self-dual five-form and asymptote to $AdS_5 \times S^5$ 
but the discussion readily generalizes to include all other 
fields or more generally to any theory with solutions 
that asymptote to $AdS_m \times X$, for some $m$ and any compact manifold
$X$. 

Let us summarize the steps involved in the construction 
of the map: 
\begin{enumerate}
\item Expand the solution (using a radial coordinate $z$ as a small parameter) 
up to certain order and write the 
deviation from $AdS_m \times X$ in terms of harmonics of $X$.
\item If the solution is not in the de Donder gauge, 
combine the fluctuations in gauge invariant variables.
\item  Use the KK map to obtain the asymptotic expansion
of the corresponding lower dimensional fields.
\item Insert the coefficients of the asymptotic expansion 
in the renormalized 1-point functions to obtain the vevs. 
 \end{enumerate}
The asymptotic expansion of the 10d solution in general will 
contain non-zero terms for many coefficients. The only ones that 
carry physical information are the ones that have the 
correct leading radial behavior to correspond to normalizable
or non-normalizable modes. The former give a contribution to 
the vev of the dual operator and the latter to the 
coefficient of the deformation of the QFT Lagrangian by the dual operator. 
We should note however that the remaining coefficients will generically
contribute to vevs of higher dimension operators
via non-linear contributions to the holographic map.

The higher the dimension of the dual operator, the higher the 
order needed in the gauge invariant variables and the 
KK map. So although this work solves the problem of computing
the vevs in principle, in practice the method becomes cumbersome
to carry out when sufficiently high 
dimension operators are involved (but since the procedure is 
algorithmic one could in principle computerize it). In this
paper we explicitly worked out the map to the first non-trivial 
order. This is sufficient to compute vevs of operators up to
dimension 4 and thus covers all relevant and marginal
operators in four dimensions. As noted in the introduction,
more efficient methods may be available when the solution
has special properties. In this paper we mainly aimed at 
settling the issue of principle in full generality.
 
To illustrate the general procedure we analyzed a solution 
that corresponds to a particular point on the Coulomb branch of 
${\cal N}=4$ SYM. This is an interesting example because the vevs
are protected by supersymmetry and therefore the supergravity 
dual must reproduce them {\it exactly}. The vevs
corresponding to fields in gauged supergravity were 
previously computed in \cite{BFS1,BFS2}. Here we computed in addition the 
vev of the operator of dimension 4 and exact agreement with 
the quantum field theory values was found! This constitutes the 
first non-trivial quantitative test of gravity/gauge theory duality away
from the fixed point that involves a vev of an operator dual
to a KK field. 

In the discussion so far we focused on how to compute vevs
starting from a given 10d solution. The (inverse of the) holographic map 
can be used to see how spacetime is reconstructed from QFT 
data. In particular, we see from our discussion that the 
vevs of operators dual to KK modes provide a harmonic resolution of the 
compact space. From a more general viewpoint, notice that 
in the radial Hamiltonian formulation of holographic 
renormalization
the vevs are associated with the radial canonical momenta conjugate to the 
sources. The holographic map therefore maps the field theory 
data to the phase space of the gravitational theory. It follows
that these data are sufficient to uniquely determine the 
bulk solution, even though the explicit formulae only provide
an asymptotic solution up to a certain order. 

One could thus ``holographically engineer'' duals of interesting 
quantum field theories by starting from the field theory vevs and 
using the holographic map. For this procedure to yield a 
smooth geometry, the vevs should clearly be large compared to
the string scale. Even if this condition is satisfied, there is still
no guarantee that a smooth geometry would emerge. For instance,
it is well known that a necessary 
condition for a {\it smooth supergravity dual} is that 
the conformal anomaly of the theory at the UV fixed point 
should satisfy $c=a$ in the large $N$ 
and $\l$ limit \cite{Henningson:1998gx}.
Let us also note that even if a given theory {\it has} a smooth dual geometry,
in practice it may not be easy to sum the 
asymptotic solution  into this smooth solution. New tools
that capture global issues of the correspondence may 
be needed to properly analyze this problem.
It would be very interesting to explore this line 
of thought further. 
   
In many cases it is clear from the construction of the 
supergravity solution (with $AdS$ asymptotics)
what the corresponding gauge theory dual is. 
For instance this is the case 
if the solution is obtained via a 
near-horizon limit from another solution.
Yet there are many other cases where solutions
with $AdS$ asymptotics have been obtained by 
directly solving the supergravity equations and 
there is no physical argument that would 
identify the dual theory.
The work presented here should be useful 
both in verifying the gauge theory duals in 
cases where a proposed identification is available and 
also for extracting the gauge theory dual 
in other cases.

\section*{Acknowledgments}

The authors are supported by NWO, KS via the Vernieuwingsimplus grant 
``Quantum gravity and particle physics'' and 
MMT via the Vidi grant ``Holography,
duality and time dependence in string theory''.

\appendix

\section{Harmonic expansion of the antisymmetric tensor} \label{5form}

We expand the antisymmetric tensor as
\bea
a_{\m \n \r \s}(x,y) &=& 
\sum \tilde{b}^{I_1}_{\m \n \r \s}(x) Y^{I_1}(y) \nonumber \\
a_{\m \n \r a}(x,y) &=& \sum (\tilde{b}^{I_5}_{(v)\m \n \r}(x) Y_a^{I_5}(y) 
+ \tilde{b}^I_{(s)\m \n \r}(x) D_a Y^{I_1}(y)) \nonumber \\
a_{\m \n a b}(x,y) &=& \sum (\tilde{b}^{I_{10}}_{(t) \m \n}(x) 
Y_{[ab]}^{I_{10}}(y) 
+ \tilde{b}^{I_5}_{(v)\m \n}(x) D_{[a} Y_{b]}^{I_5}(y)) \nonumber \\
a_{\m a b c}(x,y) &=& \sum 
(\tilde{b}^{I_5}_{(v) \m}(x) \e_{abc}{}^{de} D_d Y_e^I(y)
+ \tilde{b}^{I_{10}}_{(t)\m}(x) D_{[a} Y_{bc]}^{I_{10}}(y)) \nonumber \\
a_{a b c d}(x,y) &=& \sum (b_{(s)}^{I_1}(x) \e_{abcd}{}^{e} 
D_{e} Y^I(y) + b_{(v)}^{I_5}(x) \e_{abcd}{}^{e} Y_e^{I_5}(y))
\eea

Gauge transformations act on the 4-form as follows
\be
\d a_{MNPQ} = 4 D_{[M} c_{NPQ]} 
\ee
The antisymmetric tensor parameter has the following expansion
\bea
c_{\m \n \r}(x,y) &=& \sum c^{I_1}_{\m \n \r}(x) Y^{I_1}(y) \\
c_{\m \n a}(x,y) &=& \sum (c_{(v)\m \n}^{I_5}(x) Y_a^{I_5}(y)
+ c^{I_1}_{(s)\m \n}(x) D_a Y^{I_1}(y)) \nonumber \\
c_{\m a b}(x,y) &=& \sum (c^{I_{10}}_{(t)\m}(x) Y_{[ab]}^{I_{10}}(y) 
+ c^{I_5}_{(v)\m}(x) D_{[a} Y_{b]}^{I_5}(y)) \nonumber \\
c_{a b c}(x,y) &=& \sum (c^{I_5}_{(v)}(x) \e_{abc}{}^{de} D_d Y_e^{I_5}(y)
+ c^{I_{10}}_{(t)}(x) D_{[a} Y_{bc]}^{I_{10}}(y)) 
\eea
This implies the following gauge transformations for the fields,
\bea
\d \tilde{b}_{\m \n \r \s}^{I_1} = 4 D_{[\m} c_{\n \r \s]}^{I_1}, \qquad 
\d \tilde{b}_{(v)\m \n \r}^{I_5} = 3  D_{[\m} c_{(v) \n \r]}^{I_5}, \qquad 
\d \tilde{b}_{(s) \m \n \r}^{I_1} = - c_{\m \n \r}^{I_1} \nn \\
\d \tilde{b}_{(t)\m \n}^{I_{10}} = 2 D_{[\m} c_{(t)\n]}^{I_{10}}, \qquad
\d \tilde{b}_{(v)\m \n}^{I_5} = 
2 D_{[\m} \tilde{c}_{(v)\n]}^{I_5} + c_{(v)\m \n}^{I_5} \\
\d \tilde{b}_{(v)\m}^{I_5} = D_\m c_{(v)}^{I_5}, \qquad 
\d \tilde{b}_{(t)\m}^{I_{10}} = D_\m c_{(t)}^{I_{10}} 
- 3 c_{(t)\m}^{I_{10}}, \qquad 
\d b_{(v)}^{I_5} = -(\L^{I_5}-4) c_{(v)}^{I_5} \nn
\eea
It follows that the following combinations are gauge invariant,
\bea
b^{I_1}_{\m \n \r \s} &=& \tilde{b}^{I_1}_{\m \n \r \s} 
+ 4 D_{[\m} \tilde{b}_{(s)\n \r \s]}^{I_1} \\
b^{I_5}_{(v)\m \n \r} &=& \tilde{b}^{I_5}_{(v)\m \n \r} 
-\frac{3}{2} D_{[\m} \tilde{b}_{(v)\n \r]}^{I_5} \nn \\
b^{I_{10}}_{(t)\m \n} &=& \tilde{b}^{I_{10}}_{(t)\m \n} 
+\frac{2}{3} D_{[\m} \tilde{b}_{(t)\n]}^{I_{10}} \nn \\
b^{I_5}_{(v)\m} &=& \tilde{b}^{I_5}_{(v)\m} 
+ \frac{1}{(\L^{I_5}-4)} D_{\m} b_{(v)}^{I_5} \nn 
\eea
Indeed the field strength
\be
f_{MNPSR} = 5 D_{[M} a_{NPRS]}
\ee
can be expressed in terms of these modes.

The gauge used in \cite{Kim:1985ez},
\be
D^a a_{aMNP}=0
\ee
amounts to setting to zero
\be
\tilde{b}_{(s)\m \n \r}^{I_1}=\tilde{b}_{(v)\m \n}^{I_5}=
\tilde{b}_{(t)\m}^{I_{10}}=b_{(v)}^{I_5}=0.
\ee
Our normalizations are such that the gauge invariant 
variables evaluated 
in this gauge agree with the parametrization
in \cite{Kim:1985ez}.

\section{Spherical harmonics}

The defining equations for the spherical harmonics are
\bea
\Box_y Y^{I_1} &=& \L^{I_1} Y^{I_1}, \qquad \L^{I_1} = - k (k+4), 
\quad k=0,1,2,... \label{sc_h} \\ 
\Box_y Y_a^{I_5} &=& \L^{I_5} Y_a^{I_5}, \qquad \L^{I_5} = 
-(k^2 +4k -1), \quad k=1,2,... \nonumber \\ 
\Box_y Y_{(ab)}^{I_{14}} &=& \L^{I_{14}} Y_{(ab)}^{I_{14}}, \qquad 
\L^{I_{14}} = -(k^2 + 4k -2), \quad k=2,3,... \nonumber \\  
\Box_y Y_{[ab]}^{I_{10} } &\equiv & \L^{I_{10}}  Y_{[ab]}^{I_{10}},
\qquad  \L^{I_{10}} = -(k^2 + 4k -2), 
\quad k=1,2,... \nonumber \\ 
D^a Y_a^{I_5} &=& D^a Y_{(ab)}^{I_{14}} =D^{a} Y_{[ab]}^{I_{10} }=0. 
\nonumber 
\eea
Useful identities for the scalar harmonics include
\bea
D^{a} D_{(a} D_{b)} Y^{I} &=& 4 (1 + \frac{\Lambda^{I}}{5}) D_a Y^{I};
\\ 
\Box_y D_{(a} D_{b)} Y^{I} &=& (10 + \Lambda^{I}) D_{(a} D_{b)} Y^I; \nn
\\
\Box_y D_a Y^{I} &=& (\Lambda^{I} + 4) D_a Y^I. \nn
\eea

\subsection{Spherical harmonics with $SO(4)$ symmetry} \label{harm_ap}

We introduce the following coordinates on $S^5$
\be
ds^2= d \q^2  + \sin^2 \q  d \W_3^2 + \cos^2 \q  d \f^2.
\ee
The differential equation (\ref{sc_h}) for the scalar harmonics
is separable. Imposing $SO(4)$ symmetry implies that the spherical 
harmonics depend only on $\q$ and $\f$. The general
solution can then be expressed in terms of a hypergeometric functions,
\be
Y^{(k,m)}(\q,\f)= c_{(n,m)} y^k_m(\q) e^{i m \f}
\ee
where $c_{(n,m)}$ is a normalization constant and 
the function $y^k_m(\q)$ is given by
\be
y^k_m(x)=x^{|m|} 
{}_1F_2(-\frac{1}{2}(k-|m|),2+\frac{1}{2}(k+|m|),1 + |m|;x^2)
\ee
with $x=\cos \q$ (there are also a second solution 
with leading behavior $x^{-|m|}$ but this solution does not 
reduces to a finite polynomial for any choice of the quantum numbers). 
The hypergeometric function reduces to a finite polynomial
when either the first or second argument is zero or a negative
integer. This leads to the following cases
\be
(k=2l, \quad m=2 n), \qquad (k=2l+1, \quad m=2 n +1) 
\qquad n\in [-l,l], \ l \in Z^+ 
\ee
with
\bea 
y^{2l}_{2n}(x)&=&x^{2 |n|} {}_1F_2(-l+|n|,2+l+|n|,1 + 2 |n|;x^2) \\
y^{2l+1}_{2n+1}(x)&=& x^{|2 n+1|} {}_1F_2(-l+|n|,3+l+|n|,2 + 2 |n|;x^2) \nn
\eea
Particularly relevant for us here are harmonics 
that are also $SO(2)$ symmetric which are given by
\be \label{m0}
Y^{(2l,0)}(\q,\f) = \frac{ (-)^{l} }{2^{l} \sqrt{2l+1}}
    \left( \sum_{m=0}^{l} (-)^m 
\left ( 
\begin{array}{c} 
l \\ 
m 
\end{array} 
\right ) 
\left ( 
\begin{array}{c} 
l + m + 1 \\ 
l + 1 
\end{array} 
\right ) 
(\cos\q)^{2m} \right ).
\ee 
The lowest harmonics are therefore
\bea \label{y_expl}
Y^{(2,0)} &=& \frac{1}{2 \sqrt{3}}(3 \cos^2 \q -1), \\
Y^{(4,0)} &=& \frac{1}{4 \sqrt{5}} (10 \cos^4 \q - 8 \cos^2 \q +1), \nn \\ 
Y^{(6,0)} &=& \frac{1}{8 \sqrt{7}}
(35 \cos^5 \q - 45 \cos^4 \q + 15 \cos^2 \q -1) \nn
\eea
The overall normalization in (\ref{m0}) has been chosen so
that the harmonics are normalized as in \cite{LMRS}, i.e. 
\be \label{nor_sc}
\int Y^{I_1} Y^{I_2} = z(k) \d^{I_1 I_2}, \qquad 
z(k) = \frac{\pi^3}{2^{k-1} (k+1) (k+2)}
\ee 
Recall that the scalar harmonics can be represented as
\be \label{harm_C}
Y^{I_1} = C^{I_1}_{i_1 \cdots i_k} x^{i_1} \cdots x^{i_k}
\ee
where $x^{i_n}$ are Cartesian coordinates on $S^5$ and 
$C^I_{i_1 \cdots i_k}$ is a totally
symmetric traceless rank $k$ tensor of $SO(6)$.
The normalization in 
(\ref{nor_sc}) corresponds to delta function normalization 
for the $C^I$'s, i.e.
\be
\langle C^{I_1} C^{I_2} \rangle \equiv
C^{I_1}_{i_1 \cdots i_k}C^{I_2 i_1 \cdots i_k}
= \d^{I_1 I_2}.
\ee
For the scalar harmonics we use the following definitions:
\bea
&&\int D_{(a}  D_{b)} Y^{I_1} D_{(a} D_{b)}Y^{I_2} = 
z(k) q(k) \d^{I_1 I_2} \nn \\
&&\int Y^{I_1} Y^{I_2} Y^{I_3} = a(k_1,k_2,k_3) 
\< \cC^{I_1}\cC^{I_2}\cC^{I_3} \>
\label{3over} \\
&&\int Y^{I_1} D_a Y^{I_2} D^a Y^{I_3} = b(k_1,k_2,k_3) 
\< \cC^{I_1}\cC^{I_2}\cC^{I_3} \> \nn \\
&&\int  D^{(a}  D^{b)} Y^{I_1} D_a Y^{I_2} D_b Y^{I_3} = 
c(k_1,k_2,k_3) 
\< \cC^{I_1}\cC^{I_2}\cC^{I_3} \> \nn \\
&&\int  Y^{I_1} D^{(a} D^{b)} Y^{I_2} D_a D_b Y^{I_3} = d(k_1,k_2,k_3) 
\< \cC^{I_1}\cC^{I_2}\cC^{I_3} \> \nn \\
&&\int D^{(a}  D^{b)} Y^{I_1} (2 D_a D_c Y^{I_2} D_{(c} D_{b)} Y^{I_3}
+ D_c Y^{I_2} D_c D_{(a} D_{b)} Y^{I_3})  = e(k_1,k_2,k_3) 
\< \cC^{I_1}\cC^{I_2}\cC^{I_3} \> \nn
\eea
where 
\be \label{y_norm}
q(k) = \Lambda^{I}(4 + \frac{4}{5} \Lambda^I),
\qquad
a(k_1,k_2,k_3) =\frac{\pi^3}{(\frac{1}{2}\Sigma+2)! 2^{\frac{1}{2}(\Sigma-2)}}
\frac{k_1! k_2! k_3!}{\a_1! \a_2! \a_3!}
\ee
and $\Sigma=k_1+k_2+k_3,\ \a_1{=}\frac{1}{2}(k_2+k_3-k_1)$ etc. One 
can derive explicit formulae that express $b(k_1,k_2,k_3), c(k_1,k_2,k_3),
d(k_1,k_2,k_3), e(k_1,k_2,k_3)$ in terms of $a(k_1,k_2,k_3)$ by 
use partial integrations.
%
Useful identities include:
\bea
d(k_2,k_1,k_3) + c(k_1,k_2,k_3) + \frac{q(k_1)}{\Lambda^{I_1}}
b(k_2,k_1,k_3) &=& 0; \nn \\
b(k_2,k_1,k_3) + b(k_1,k_2,k_3) + \Lambda^{I_3} a(k_1,k_2,k_3) &=& 0 \nn.
\eea 
We further have
\be
D_{(\q} D_{\q)} Y_0^2 = \frac{6}{5}(1 - 2 \cos 2 \q), \qquad
D_{(\q} D_{\q)} Y_0^4 = \frac{12}{5}(2 + \cos 2 \q -5 \cos 4 \q). 
\ee
For the tensor harmonics we use the following definitions
\bea
&&\int Y_{ab}^{I_1} D_{a}Y^{I_2} D_{b} Y^{I_3} =
c^{(t)}(k_1^{(t)},k_2,k_3); \\
&&\int Y_{ab}^{I_1} (2 D_a D_c Y^{I_2} D_{(c} D_{b)} Y^{I_3}
+ D_c Y^{I_2} D_c D_{(a} D_{b)} Y^{I_3}) =
e^{(t)} (k_1^{(t)},k_2,k_3). \nn
\eea
The normalization of the spherical harmonic is defined as
\be
\int Y_{ab}^{I_1} Y_{ab}^{I_2} = z_{(t)} (k^{(t)}) \d^{I_1 I_2}.
\ee
The only tensor harmonic of relevance here is
\be \label{t_harm}
Y_{\q \q}^{(2,0)} = -3, \qquad 
Y_{\f \f}^{(2,0)}=\cos^2 \q (-3 + 15 \cos^2 \q), \qquad
Y_{\psi^a \psi^a}^{(2,0)}=\sin^2 \q (2 - 5 \cos^2 \q) \nn
\ee
where $\psi^a$ are the coordinates on $S^3$; we thus do not need to
discuss the tensor harmonics more generally. 

\section{Field equations up to second order} \label{2ndorder}

We discuss in this appendix 
the derivation of the field equations up to second order in fluctuations.
 
The linearized equations read
\bea
E_{MN}^{(1)} &\equiv&  R_{MN}^{(1)} 
+ \frac{4}{3!} h^{KL} F^o_{MKM_1M_2M_3} F^o_{NL}{}^{M_1M_2M_3} \\
&-&\frac{1}{3!}(f_{MM_1M_2M_3M_4} F^o_N{}^{M_1M_2M_3M_4}
+ f_{NM_1M_2M_3M_4} F^o_M{}^{M_1M_2M_3M_4}) = 0 \nn \\
E^{(1)}_{M_1\ldots M_5} &\equiv&
(f-f^*)_{M_1\ldots M_5} + \frac{1}{2} h^L_L F^o_{M_1\ldots M_5} 
-5 h^K_{[M_1} F^o_{M_2\ldots M_5]K} = 0
\eea
where 
\be
R_{MN}^{(1)} = D_K h^K_{MN} - \frac{1}{2} D_M D_N h^L_L, \qquad
h^K_{MN} = \frac{1}{2} (D_M h^K_N +D_N h_M^K - D^K h_{MN}). 
\ee
Projecting these equations onto the various harmonics leads to the 
following equations\footnote{In comparing with \cite{Kim:1985ez}
one should note that we expand in harmonics $h_{\m \n}$ rather 
than $h_{\m \n}'$ (compare (2.5)-(2.7) with our (\ref{fluct_h})).} 
\bea
(E^{(1)}_{a b})|_{D_{(a} D_{b)} Y^{I_1}} =0
&\Rightarrow &
(\frac{1}{2} \hat{h}^{\s}_{\s}{}^{I_1} + \frac{3}{10} \hat{\pi}^{I_1}) =0; 
\label{linear}\\
(E^{(1)}_{a b})|_{Y_{ab}^{I_{14}}} =0 &\Rightarrow & 
\frac{1}{2} 
(\Box + \L^{I_{14}} -2) \hat{\f}_{(t)}^{I_{14}}=0; 
\nn \\ 
(E^{(1)}{}_{a}^{\; a})|_{Y^{I_1}} &\Rightarrow & \frac{1}{10}(
(\Box+\L^{I_1}-32) \hat{\pi}^{I_1} + 80 \L^{I_1} \hat{b}^{I_1} +
\L^{I_1} (\hat{h}^{\s}_{\s}{}^{I_1} + \frac{3}{5} \hat{\pi}^{I_1}))=0; 
\nn \\
(E^{(1)}_{\m\n\r\s a})|_{ D_a Y^{I_1}}=0
&\Rightarrow &(\hat{b}^{I_1}_{\m \n \r \s} 
+ \e_{\m \n \r \s}{}^\t D_\t \hat{b}^{I_1}) =0; \nn \\
(E^{(1)}_{\m\n\r\s\t})|_{Y^{I_1}}=0
&\Rightarrow &(5 D_{[\m} \hat{b}_{\n \r \s \t]}^{I_1} - \e_{\m \n \r \s}{}^\t
(\frac{1}{2} \hat{h}^{\s I_1}_\s + \L^{I_1} \hat{b}^{I_1} 
- \frac{1}{2} \hat{\pi}^{I_1}))=0. \nn
\eea
These equations lead to the scalar field equations quoted in section 
\ref{lin_ord} upon elimination of $\hat{b}_{\m \n \r \s}$ and 
$\hat{h}^{\s}_{\s}$ and then diagonalization.

We now move to the quadratic order. The field equations are 
\be
E_{MN}^{(1)} = T_{MN}^{(2)}, \qquad 
E^{(1)}_{M_1 \ldots M_5} = T^{(2)}_{M_1 \ldots M_5}
\ee
where the quadratic corrections are given by \cite{Arutyunov:1999en}
\bea
&&T^{(2)}_{M_1 \ldots M_5} = -\frac{1}{2} h^L_L f^*_{M_1\ldots M_5} 
+5 h^K_{[M_1} f^*_{M_2\ldots M_5]K} \la{cor1} \\
&&-\frac{5}{2} h^L_L  h^K_{[M_1} F^o_{M_2\ldots M_5]K} 
+(\frac{1}{8} (h^L_L)^2 +\frac{1}{4} h^{ML} h_{ML}) F^o_{M_1 \ldots M_5}
+10 h^{K_1}_{[M_1} h^{K_2}_{M_2} F^o_{M_3M_4M_5]K_1K_2} \nn \\
&&T_{MN}^{(2)} = -R_{MN}^{(2)} \la{cor2} \\
&&+ \frac{4}{3!} h^{KL} h_L^S F^o_{MNM_1M_2M_3} F^o_{NS}{}^{M_1M_2M_3}
+\frac{2 \cdot 3}{3!} h^{K_1S_1} h^{K_2S_2} 
F^o_{MK_1K_2M_1M_2} F^o_{NS_1S_2}{}^{M_1M_2} \nn \\ 
&&-\frac{4}{3!} h^{KS} (f_{MKM_1M_2M_3} F^o_{NS}{}^{M_1M_2M_3}+
f_{NKM_1M_2M_3} F^o_{MS}{}^{M_1M_2M_3})
+\frac{1}{3!} f_{MM_1\ldots M_4} f_{N}{}^{M_1\ldots M_4} \nn
\eea
where 
\be
R_{MN}^{(2)}= - D_K(h_L^K h^L_{MN}) + \frac{1}{2} D_N(h_{KL} D_M h^{KL})
+\frac{1}{2} h_{MN}^K D_K h^L_L - h_{MK}^L h_{NL}^K
\ee
These quantities were computed to second order in the field $s$ in the
de Donder gauge in \cite{LMRS}, by substituting the linear solution of
the field equations
\bea
&& h^{I_1}_{\m \n} = U(k) {s}^{I_1} g^{o}_{\m\n} + W(k) D_{(\m} D_{\n)}
s^{I_1}; \\
&& h^{I_1}_{\a \b} = V(k) s^{I_1} g_{\a\b}, \qquad b^{I_1} = - s^{I_1}; 
\nn \\
&& V(k) = - \frac{5}{3} U(k) = 2k, \qquad W(k) = \frac{4}{k+1}. \nn
\eea
As discussed in the main text, the resulting field equations will be
applicable to other gauge choices provided that one replaces each
field by the corresponding gauge invariant field. In particular, for
the field  ``$s^{I_1}$'' must denote the appropriate gauge invariant field. 
For computing the quadratic corrections, however, it is sufficient to
use the field $\hat{s}^{I_1}$ which is gauge invariant to linear
order, since the difference between this field and the gauge invariant
field is itself quadratic in fluctuations. 

\subsection{Scalar fields}

To compute the corrected scalar equations 
we will need to use the following components of (\ref{cor1}) and
(\ref{cor2}):
\bea
Q_1^1 & \equiv & \frac{1}{5} (T^{(2)}_{a b})|_{D_{(a} D_{b)} Y^1}; \\
&=& \frac{1}{20 q_1 z_1} \left ( (c_{123} + d_{231} + d_{321}) T_{23}
+ 32 c_{123} D_{\m} \hat{s}^2 D^{\m} \hat{s}^3 \right ), \nn \\
T_{23} &=& ( 3 V_2 V_3 + 5 U_2 U_3) \hat{s}^2 \hat{s}^3 + W_2 W_3
D^{(\m} D^{\n)} \hat{s}^2 D_{(\m} D_{\n)} \hat{s}^3. \nn
\eea
The notation in the first line implies the projection of the tensor
(which is quadratic in spherical harmonics) onto the spherical
harmonic. (Note that the factor of five is included in the definition
so as to match the conventions of \cite{LMRS}). Similarly one has
\bea
Q_2^1 & \equiv & \frac{1}{5} (T^{(2)}_{a}{}^{\; a})|_{Y^1}
= \frac{1}{20 z_1} \left ( 10 S_{123} + T_{23}(b_{123} - 2 f_3
a_{123}) + 32 D_{\m} \hat{s}^{2} D^{\m} \hat{s}^2 b_{123} \right ),
\nn \\
S_{123} &=& V_3 U_2 a_{123}
D^{\m} (\hat{s}^2 D_{\m} \hat{s}^3) + W_2 V_3 a_{123} 
D_{\m} (D^{(\m} D^{\n)} \hat{s}^2 D_{\n} \hat{s}^2) \\
&& - V_2 V_3 b_{213} \hat{s}^2 \hat{s}^3
- 8 (a_{123} f_2 f_3 \hat{s}^2 \hat{s}^3 + b_{123} D^{\m}\hat{s}^2
D_{\n} \hat{s}^3) - a_{123} V_2 (64 f_3 + 80 V_3) \hat{s}^2
\hat{s}^3. \nn 
\eea
We also need
\bea
(T^{(2)}_{\m\n\r\s a})|_{ D_a Y^1} & \equiv & - \ep_{\m\n\r\s\t}
Q_3^{\t 1}, \\
Q_{3}^{\t 1} &=& - \frac{1} {f_1 z_2} \left ( (U_2 + 3 V_2)\hat{s}^2 
D^{\t} \hat{s}^3 + W_2 D^{(\t} D^{\r )} \hat{s}^2 D_{\r} \hat{s}^3
\right ) b_{213}; \nn \\
(T^{(2)}_{\m\n\r\s\t})|_{Y^1} & \equiv & Q_{4}^1 \ep_{\m\n\r\s\t}, \\
Q_{4}^1 &=& - \frac{1}{4 z_1} \left ( T_{23} - (16 V_2 f_3 + 40 V_2
V_3) \hat{s}^2 \hat{s}^3 \right ) a_{123}. \nn
\eea
For the scalar field coupling to the tensor harmonic we need
\bea
Q_{(t)}^{I_{14}} & \equiv & (T^{(2)}_{a b})|_{Y_{ab}^{I_{14}}}, \\
&=& \frac{1}{4 z_{(t)1}} \left ( d^{(t)}_{123} T_{23} + 32
c^{(t)}_{123} D_{\m} \hat{s}^2 D^{\m} \hat{s}^3 \right ). \nn
\eea

Then the corrected equations of motion are written in terms of
quantities just defined as
\bea
(\Box - k(k-4)) s^{I_1} &=& \frac{1}{2 (k+2)} \left ( (k+4)(k+5) Q_1 + Q_2
+ (k+4) (D_{\m}Q_3^{\m} + Q_4) \right )^{I_1}; \nn \\
(\Box - (k+8)k(k+4)) t^{I_1} &=& \frac{1}{2 (k+2)} \left ( k(k-1) Q_1 + Q_2
- k (D_{\m}Q_3^{\m} + Q_4) \right )^{I_1}, 
\eea
whilst the corrected equation for the scalar $\f_{(t)}^{I_{14}}$ is
\be
(\Box - k(k+4)) \f_{(t)}^{I_{14}} = 2 Q_{(t)}^{I_{14}}.
\ee

\subsection{Tensor fields}

To compute the correction to the metric and KK gravitons we 
we need $T^{(2)}_{\m\n}$. For brevity we will include only
the terms of interest here, namely $\hat{s}^2$. The curvature
contribution to $(T^{(2)}_{\m\n})|_{Y^{I}}$ is then
\bea
&& \frac{1}{z_{I}} \left ( - \frac{4}{9} a_{I22} D_{\m} D_{\r} D_{\s} \hat{s}^2
D_{\n} D^{\r} D^{\s} \hat{s}^2  - \frac{32}{3} a_{I22} D^{\r} \hat{s}^2 D_{\r}
D_{\m} D_{\n} \hat{s}^2 - \frac{8}{9} b_{I22} D_{\m} D_{\r} \hat{s}^2
D_{\n} D^{\r} \hat{s}^2 \right . \nn \\
&& \qquad + \frac{8}{9} a_{I22} 
(D^{\r} D^{\s} \hat{s}^2 D_{\r} D_{\s} \hat{s}^2)
g^{o}_{\m\n} + (\frac{40}{9} b_{I22} - 32  a_{I22}) \hat{s}^2 
D_{\m} D_{\n} \hat{s}^2 \\
&& \left. \qquad - \frac{136}{9} a_{I22} (D_{\m} \hat{s}^2 D^{\n} \hat{s}^2)
+ \frac{32}{9} (a_{I22} - b_{I22}) (\hat{s}^2)^2 g^{o}_{\m\n} \right)\nn
\eea
whilst the field strength contribution to $(T^{(2)}_{\m\n})|_{Y^{I}}$ is
\bea
&& \frac{1}{z_I} \left ( g^{o}_{\m\n} (\frac{32}{9} a_{I22}(2 (\hat{s}^2)^2 -
 D_{\r} D_{\s} \hat{s}^2 D^{\r} D^{\s} \hat{s}^2) - 4
b_{I22} (D_{\r} \hat{s}^2 D^{\r} \hat{s}^2) ) 
- \frac{64}{3} a_{I22} \hat{s}^2 D_{\m} D_{\n} \hat{s}^2 \right . \nn \\
&&  \left .
\qquad + 8 D_{\m} \hat{s}^2 D_{\n} \hat{s}^2 b_{I22} \right ). 
\eea
These lead to the following equation for the graviton,
\be
(L_E + 4) h_{\m \n}^0 = T_{\m \n}^{(2)}|_{Y^0} - g^o_{\m \n} 
(\frac{5}{3} Q_2^0 + 8 Q_4^0)|_{Y^0}
\ee
where the linearized Einstein operator is defined as usual by
\be \label{lin_ein}
L_{E} \l_{\m\n} = \frac{1}{2} ( - \Box \l_{\m\n} + D_{\r} D_{\m}
\l^{\r}_{\; \n} + D_{\r} D_{\n} 
\l^{\r}_{\; \m} - D_{\m} D_{\n} \l^{\r}_{\r} ).
\ee
The term proportional to $Q_2$ and $Q_4$ arise when eliminating 
$\Box \tilde{\pi}^0$ and $\tilde{h}^{\s 0}_\s$ from the equation.

The following identities prove useful:
\bea
L_E ((D^\r D^\s \hat{s}^2 D_\r D_\s \hat{s}^2)g_{\m \n}^o) &=&
-\frac{1}{2} \Box (D^\r D^\s \hat{s}^2 D_\r D_\s \hat{s}^2)g_{\m \n}^o
-3 D_\m D_\n D_\r D_\s \hat{s}^2  D^\r D^\s \hat{s}^2 \nn \\
&& -3 D_\m D_\r D_\s \hat{s}^2  D_\n D^\r D^\s \hat{s}^2 \nn \\
L_{E} (D_{\m} D_{\r} \hat{s}^2 D_{\n} D^{\r} \hat{s}^2) 
&=& 2 D^{\r} D_{\m} \hat{s}^2 D_{\r} D_{\n} \hat{s}^2 - D_{\m} D^{\r}
D^{\s} \hat{s}^2 D_{\n} D_{\r} D_{\s} \hat{s}^2 \nn \\
&&+ g^{o}_{\m\n} (D^{\r}D^{\s} \hat{s}^2 D_{\r} D_{\s} \hat{s}^2) 
- 9 D^{\r} \hat{s}^2 D_{\m}
D_{\n} D_{\r} \hat{s}^2 - 7 D_{\m} \hat{s}^2 D_{\n}\hat{s}^2  \nn \\
&&+ 12 \hat{s}^2 D_{\m} D_{\n} \hat{s}^2 - (D^{\r} \hat{s}^2 D_{\r}
\hat{s}^2) g^{o}_{\m\n}, \label{lact} \\
L_{E} ( (D^{\r} \hat{s}^2 D_{\r}
\hat{s}^2) g^{o}_{\m\n}) &=& 
8 (D^{\r} \hat{s}^2 D_{\r}
\hat{s}^2) g^{o}_{\m\n} - g^{o}_{\m\n} (D^{\r}
D^{\s} \hat{s}^2 D_{\r} D_{\s} \hat{s}^2) \nn \\
&& 
- 3 D^{\r} \hat{s}^2 D_{\m} D_{\n} D_{\r} \hat{s}^2 
- 3 D^{\r}
D_{\m} \hat{s}^2 D_{\r} D_{\n} \hat{s}^2; \nn \\
L_{E} (\hat{s}^2 D_{\m} D_{\n}\hat{s}^2) &=& 
- 3 D_{\m} \hat{s}^2 D_{\n}\hat{s}^2 
- (D^{\r} \hat{s}^2 D_{\r}\hat{s}^2) g^{o}_{\m\n} 
+ D_{\m} D^{\r} \hat{s}^2 D_{\n} D_{\r} \hat{s}^2; \nn \\
L_{E} ( (\hat{s}^2)^2 g^{o}_{\m \n}) &=& 4 (\hat{s}^2)^2 g^{o}_{\m \n} -
(D^{\r} \hat{s}^2 D_{\r} \hat{s}^2) g^{o}_{\m\n} -
3 \hat{s}^2 D_{\m} D_{\n} \hat{s}^2 - 3 D_{\m} \hat{s}^2
D_{\n}\hat{s}^2. \nn
\eea

\end{document}